\documentclass[iop]{emulateapj}
\usepackage{apjfonts}
\usepackage{graphicx}

\def\d3{$\delta_{3}$ }
\def\1d3{$(1 + \delta_{3})$ }
\def\l1d3{$\log_{10}(1 + \delta_{3})$ }

\def\s3{$\Sigma_{3}$}

\def\24m{24 $\mu$m}

\def\sm{$M_{*}$}

\def\kms{${\rm km~s^{-1}}$ }

\def\Msolar{$\rm M_{\odot}$}

\def\photoz{photo-$z$}

\def\deltaz{$\sigma_{\Delta~z/(1+z_{s})}$}
\def\outlier{$\eta$}
\def\fq{$f_{q}$}
\newcommand{\gps}{\ensuremath{g_{\rm P1}}}
\newcommand{\rps}{\ensuremath{r_{\rm P1}}}
\newcommand{\ips}{\ensuremath{i_{\rm P1}}}
\newcommand{\zps}{\ensuremath{z_{\rm P1}}}
\newcommand{\yps}{\ensuremath{y_{\rm P1}}}
\newcommand{\wps}{\ensuremath{w_{\rm P1}}}
\newcommand{\grizy}{\gps\rps\ips\zps\yps}
\newcommand{\PS}{\protect \hbox {Pan-STARRS1}}

\shorttitle{The role of group environment in the SFR$-$\sm relation in PS1}
\shortauthors{Lin et al.}

\begin{document}

\title{The Pan-STARRS1 Medium-Deep Survey: The role of galaxy group environment in the star formation rate versus stellar mass relation and quiescent fraction out to $z \sim 0.8$}

\author{Lihwai Lin \altaffilmark{1}, Hung-Yu Jian \altaffilmark{2}, Sebastien Foucaud \altaffilmark{3,1}, Peder Norberg \altaffilmark{4}, R. G. Bower \altaffilmark{4}, Shaun Cole \altaffilmark{4}, Pablo Arnalte-Mur \altaffilmark{4}, Chin-Wei Chen \altaffilmark{1}, Jean Coupon \altaffilmark{1,11}, Bau-Ching Hsieh \altaffilmark{1}, Sebastien Heinis \altaffilmark{5}, Stefanie Phleps \altaffilmark{6},  Wen-Ping Chen \altaffilmark{7}, Chien-Hsiu Lee \altaffilmark{8,6}, William Burgett \altaffilmark{9}, K. C. Chambers\altaffilmark{9}, L. Denneau\altaffilmark{9}, P. Draper\altaffilmark{4}, H. Flewelling\altaffilmark{9}, K. W. Hodapp\altaffilmark{9}, M. E. Huber\altaffilmark{9}, N. Kaiser\altaffilmark{9}, R.-P. Kudritzki\altaffilmark{9}, E. A. Magnier\altaffilmark{9}, N. Metcalfe\altaffilmark{4}, Paul A. Price\altaffilmark{10}, J. L. Tonry\altaffilmark{9}, R. J. Wainscoat\altaffilmark{9}, C. Waters\altaffilmark{9}}

\altaffiltext{1}{Institute of Astronomy \& Astrophysics, Academia Sinica, Taipei 106, Taiwan   (R.O.C.); Email: lihwailin@asiaa.sinica.edu.tw}
\altaffiltext{2}{Department of Physics, National Taiwan University, Taipei, Taiwan (R.O.C.)}
\altaffiltext{3}{Department of Earth Sciences, National Taiwan Normal University, N$^{\circ}$88, Tingzhou Road, Sec. 4, Taipei 11677, Taiwan (R.O.C.)}
\altaffiltext{4}{Institute for Computational Cosmology, Department of Physics, Durham University, South Road, Durham DH1 3LE, UK}
\altaffiltext{5}{Department of Astronomy, University of Maryland, Maryland, US}  
\altaffiltext{6}{Max-Planck-Institut f\"ur Extraterrestrische Physik,
Giessenbachstra{\ss}e, D-85748 Garching, Germany}
\altaffiltext{7}{Graduate Institute of Astronomy, National Central University, Chung-Li 32054, Taiwan (R.O.C.)}
\altaffiltext{8}{University Observatory Munich, Scheinerstrasse 1, 81679 Munich, Germany}
\altaffiltext{9}{Institute for Astronomy, University of Hawaii, 2680 Woodlawn Drive, Honolulu HI 96822}
\altaffiltext{10}{Department of Astrophysical Sciences, Princeton University, Princeton, NJ 08544, USA}
\altaffiltext{11}{Department of Astronomy, University of Geneva, ch. d’Ecogia 16,
CH-1290 Versoix, Switzerland}

\begin{abstract}

Using a large optically-selected sample of field and group galaxies drawn from the Pan-STARRS1 Medium-Deep Survey (PS1/MDS), we present a detailed analysis of the specific star formation rate (SSFR) -- stellar mass (\sm) relation, as well as the quiescent fraction versus \sm~ relation in different environments. While both the SSFR and the quiescent fraction depend strongly on stellar mass, the environment also plays an important role. Using this large galaxy sample, we confirm that the fraction of quiescent galaxies is strongly dependent on environment at a fixed stellar mass, but that the amplitude and the slope of the star-forming sequence is similar between the field and groups: in other words, the SSFR-density relation at a fixed stellar mass is primarily driven by the change in the star-forming and quiescent fractions between different environments rather than a global suppression in the star formation rate for the star-forming population. 
However, when we restrict our sample to the cluster-scale environments ($M>10^{14}$\Msolar), we find a global reduction in the SSFR of the star forming sequence of  17\% at 4$\sigma$ confidence as opposed to its field counterpart. After removing the stellar mass dependence of the quiescent fraction seen in field galaxies, the excess in the quiescent fraction due to the environment quenching in groups and clusters is found to increase with stellar mass, although deeper and larger data from the full PS1/MDS will be required to draw firm conclusions. We argue that these results are in favor of galaxy mergers to be the primary environment quenching mechanism operating in galaxy groups whereas strangulation is able to reproduce the observed trend in the environment quenching efficiency and stellar mass relation seen in clusters. Our results also suggest that the relative importance between mass quenching and environment quenching depends on stellar mass -- the mass quenching plays a dominant role in producing quiescent galaxies for more massive galaxies, while less massive galaxies are quenched mostly through the environmental effect, with the transition mass around $1-2\times10^{10}$\Msolar~ in the group/cluster environment.

\end{abstract}

\keywords{galaxies:clustering $-$ galaxies:evolution $-$ galaxies: high-redshift $-$ large-scale
structure of Universe}

\section{INTRODUCTION}

One of the most important findings related to galaxy formation and evolution in the last three decades is the well-known correlation between galaxy properties and their hosting environments, often referred as the star formation rate (hereafter SFR)--density, morphology--density, and color--density relations \citep{dre80,bal98,gom03,kau04,ger07,coo07,elb07}. On average, galaxies tend to be more massive, redder, and less active in star formation in denser environments, such as galaxy groups and clusters. The main drivers for this relationship can be divided into two categories: the consequence of an older population in massive halos due to the earlier formation of galactic haloes in overdense regions that are destined to become galaxy clusters, the so-called ``nature'' process, versus ``nurture'' process, which refers to the physical mechanisms acting on the galaxies located in groups or clusters. In the ``nature'' processes, galaxies living in more massive halos are formed earlier and have accumulated more stellar masses compared to galaxies in the field. As a result, galaxies residing in denser environments show older stellar populations and their stellar mass (luminosity) function is shifted towards the more massive (luminous) end \citep{kau04,bal06,rob06,muz12}. Examples of the ``nurture'' processes include ram-pressure stripping \citep{gun72,dre83}, high speed galaxy encounters \citep[galaxy harassment;][]{moo96}, galaxy-galaxy mergers \citep{mih94}, and removal of warm and hot gas \citep[strangulation;][]{lar80,bal00}. Undoubtedly both the processes of nature and nurture are responsible for the observed SFR--density relation, however, the key question is which one plays a more important role in the evolution of galaxies.  

One intriguing way to decipher the relative role between nature and nurture processes is to compare the SFR of galaxies for a given stellar mass (hereafter \sm). The nature process predicts that the main difference in galaxy properties in different environments is driven by the difference in their stellar mass distributions. In contrast, the nurture process alters the properties of galaxies of a given stellar mass. The star formation rate--stellar mass (SFR--$M_*$) relation therefore provides key insights into the physical processes that drive and regulate the star formation activities in galaxies as well as the stellar mass assembly histories of galaxies. 

A tight correlation between galaxy SFR and \sm~ for star-forming galaxies (the so-called ``main sequence'') has been observed for star-forming galaxies out to $z \sim 2$ \citep{bri04,noe07,elb07,dad07,pan09,mag10,lin12,whi12,hei13}. The normalization, slope, and scatter of this SFR--$M_*$ relation carry a wealth information on the evolution of galaxies. For example, earlier works have found that the amplitude of this main sequence increases from $z \sim 0$ to $z \sim 2$, which can be understood as the global change of the gas density over time, being greater at higher redshifts. On the other hand, the slope of this relationship directly probes the star formation rate efficiency as a function of stellar mass: slope smaller than unity means that on average less massive galaxies form stars with greater efficiency. Finally, the scatter of the main sequence is found to remain approximately constant over this redshift range \citep{dad07,whi12}, indicating that the physical processes leading to a smooth and steady supply of gas at higher redshifts are not that different from the ones which act at lower redshifts. 

To better understand the role of environment in shaping galaxy properties, it is therefore intriguing to look into the environment dependence of the SFR--$M_*$ relation,  in addition to its redshift evolution. Recent studies using local samples have shown that the environment mostly changes the fraction of passive galaxies, resulting in the observed SFR--density relation, but has little effect on the SFR--$M_*$ relation of the star-forming galaxies \citep{pen10,wij12}. This implies that the timescale of the quenching process occurring in dense environment must be relatively short so that galaxies move quickly from the star-forming sequence into the passive population, without changing the mean properties of the main-sequence. One favorable scenario responsible for environment quenching is attributed to the so-called ``satellite quenching'': galaxies experience truncation of their star formation due to tidal stripping, ram-pressure stripping, and/or shock heating when they fall into bigger halos \citep{mcg11,bol10,wet11}. 

Extending this kind of study beyond local Universe is challenging because of the difficulties in acquiring a large number of stellar mass selected samples with spectroscopic redshifts that are ideal for the environment measurement or identification. Several attempts have been made to push the study on the environment effect using group and cluster samples out to $z\sim2$ \citep{pat11,vul10,muz12,koy13} but the conclusions are still controversial. Most recently, \citet{koy13} compared the SFR--$M_*$ relation for the H$_{\alpha}$-selected sample between field and clusters out to $z \sim 2.2$ and found that the SFR--$M_*$ relation in clusters evolves in a similar manner as in the field since $z \sim 2$. However, their sample size is still small (one cluster per redshift bin), and they were not able to quantify the difference in the fraction of passive (quiescent) populations between different environments due to the H$_{\alpha}$-selection. 

In this work, we take advantage of the large sample of galaxy groups and field galaxies from the early Pan-STARRS (short for the Panoramic Survey Telescope \& Rapid Response System) Medium Deep Survey to quantify the differences in galaxy properties between the field and group environments. More specifically, we study the SFR--$M_*$~ distribution as well as the quiescent fraction vs stellar mass relation as a function of environment and redshift in the redshift interval $0.2<z<0.8$. Our goal is to understand whether the SFR--density relation is purely due to the higher proportion of quiescent galaxies in groups, or whether it is driven by the suppression of star formation in galaxies of all stellar mass in the group environment. The large survey volume enclosed in this data also allows us to divide the groups sample into group and cluster environments so that one can gain better insights on how different physical structures affect the galaxy properties. To achieve our goals, we introduce a scheme to properly correct for the field contamination and incompleteness of the group members to uncover the underlying SFR and \sm~ distribution of group galaxies (see \S\ref{sec:sfrsm} and Appendix \ref{app:corr}). Using this approach, we are allowed to construct the largest sample up-to-date for the study of the SFR--\sm~ relation in different environments at intermediate redshifts.

Our paper is structured as follows. In \S2, we describe the data and methods of measuring redshift, SFR, \sm~ and group identification used in this analysis. We present the main results in \S3. \S4 discusses the important implications for our results in understanding the evolution of  galaxies. We present our conclusions in \S5. Throughout this paper we adopt the following cosmology: \textit{H}$_0$ = 100$h$~\kms Mpc$^{-1}$, $\Omega_{\rm m} =
0.3$ and $\Omega_{\Lambda } = 0.7$. We adopt the Hubble constant $h$ = 0.7 when calculating rest-frame
magnitudes. We use a Salpeter IMF when deriving stellar masses and star formation rates. All magnitudes are given in the AB system.

\begin{figure*}[h]
\includegraphics[angle=-270,width=19cm]{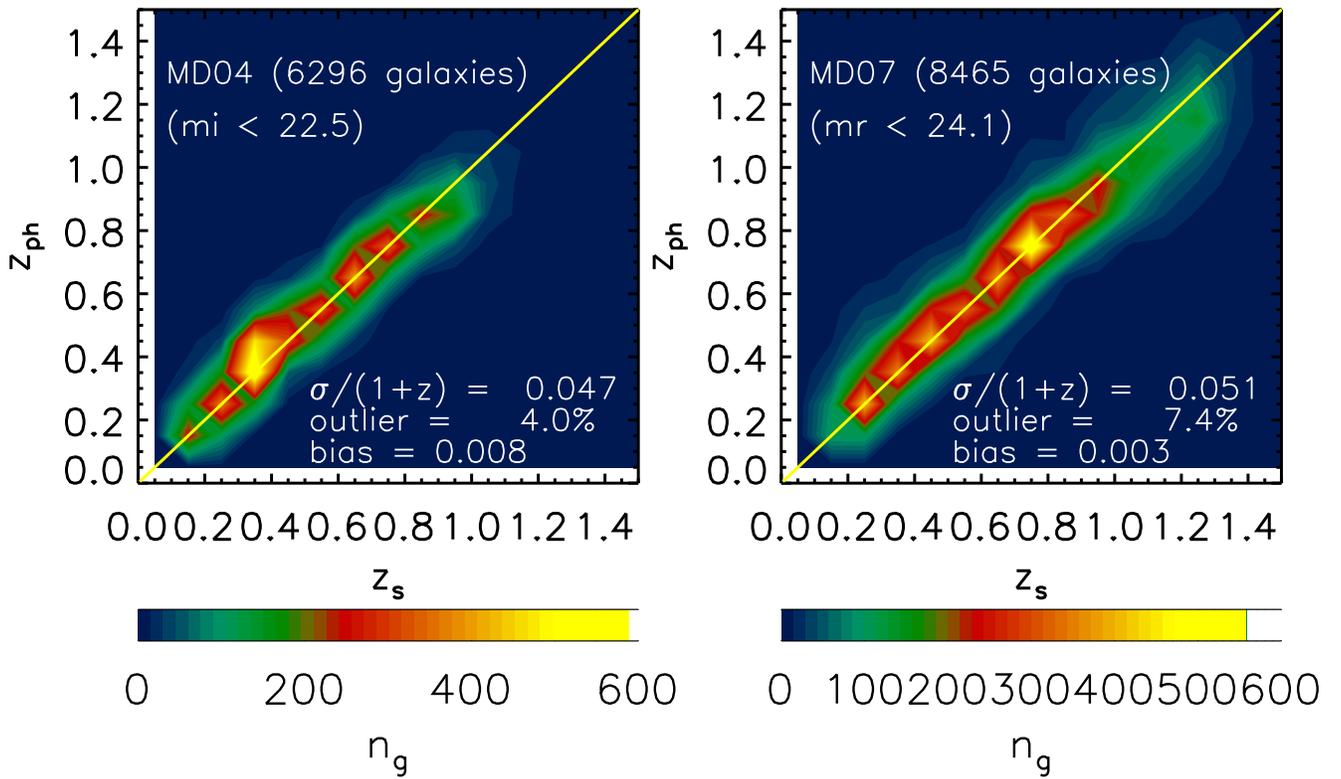}
\caption{Photometric redshifts computed using PS1MD $g_{\mathrm{P}1}r_{\mathrm{P}1}i_{\mathrm{P}1}z_{\mathrm{P}1}y_{\mathrm{P}1}$ and CFHT $u^{*}$ filters versus spectroscopic redshifts available in MD04 (left) and MD07 (right). The spectroscopic redshifts in MD04 are taken from the zCOSMOS 10K sample \citep{lil07}, and those in MD07 are drawn from the DEEP2 survey \citep{new13}. Outliers are defined as objects with $\Delta~z > 0.15\times(1+z_{s})$.
\label{fig:photoz}}
\end{figure*}

\begin{figure}
  \centering
  \includegraphics{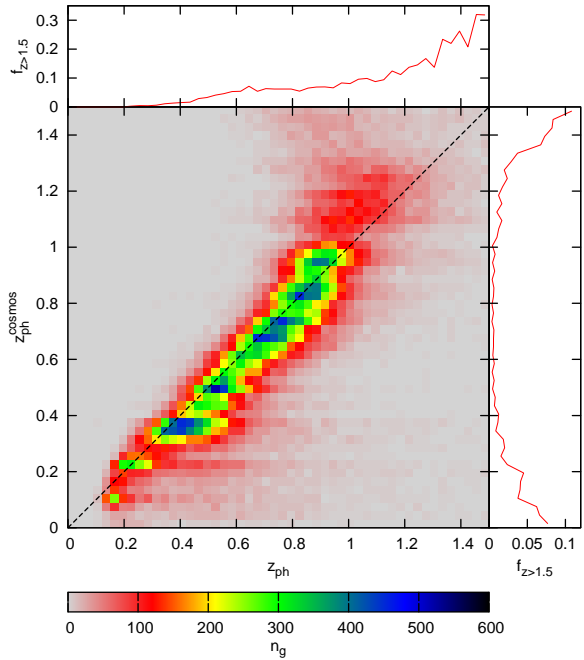}
  \caption{Photometric redshifts computed using the CFHT $u*$ and PS1MD $g_{\mathrm{P}1}r_{\mathrm{P}1}i_{\mathrm{P}1}z_{\mathrm{P}1}y_{\mathrm{P}1}$ filters versus the Ilbert et al. (2010) COSMOS photometric redshifts computed using 31 filters for the $100194$ matched objects with $i_{p1} < 24$. The colour map shows the number of galaxies ($n_g$) in bins of $0.03\times0.03$. The top graph shows the fraction of objects with a given value of $z_{ph}$ which have $z_{ph}^{cosmos} > 1.5$, and conversely the right-side graph shows the fraction of objects with a given value of $z_{ph}^{cosmos}$ and $z_{ph} > 1.5$.}
  \label{fig:photoz2}
\end{figure}

\begin{figure}[h]
\includegraphics[angle=-270,width=14cm]{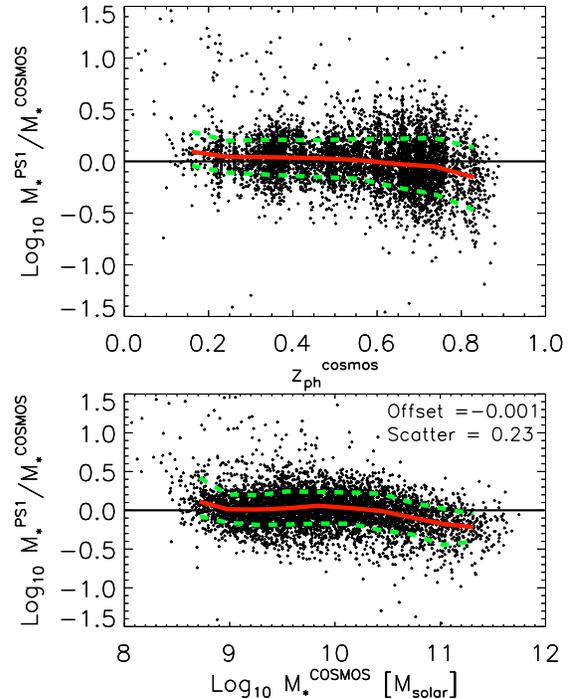}
\caption{The difference in the stellar mass between our measurement and that from the COSMOS photoz v1.7 catalog \citep{ilb10} after correcting for the difference in IMF, as a function of redshift (upper panel) and stellar mass (lower panel) for a matched sample in the COSMOS field (MD04). The black points are for all matched galaxies with whose PS1 \photoz~ agrees with the COSMOS \photoz~(i.e, ($z_{ph}^{ps1} - z_{ph}^{cosmos}$)/(1 + $z_{ph}^{cosmos}$) < 0.05). The red curve shows the median of the offset, while the lower and upper green curves represent the 16th and 84th percentiles respectively. The scatter in this comparison is $\sim$ 0.23 dex. \label{fig:sm}}
\end{figure}

\begin{figure}[h]
\includegraphics[angle=-270,width=14cm]{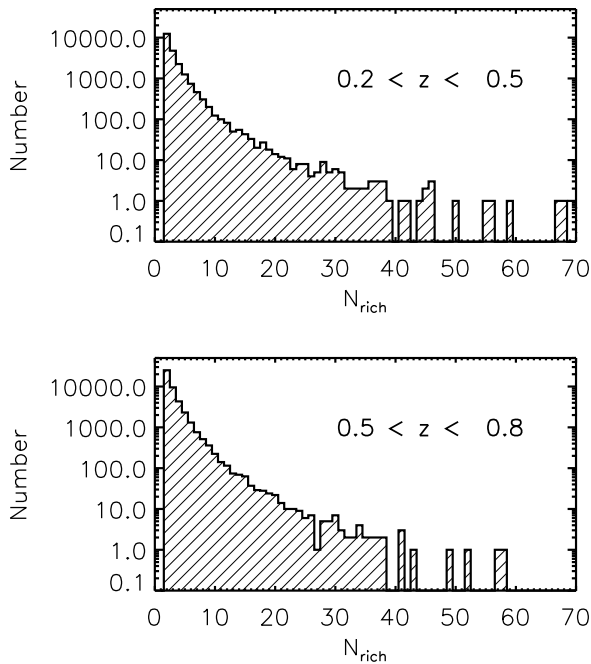}
\caption{Richness ($N_{rich}$) distributions of the PFOF groups identified in MD04 and MD07. The distributions for the group samples in 0.2 < z < 0.5 and 0.5 < z < 0.8 are shown in the top and bottom panels, respectively. \label{fig:grouphist}}
\end{figure}

\begin{figure*}[h]

\includegraphics[angle=-270,width=19cm]{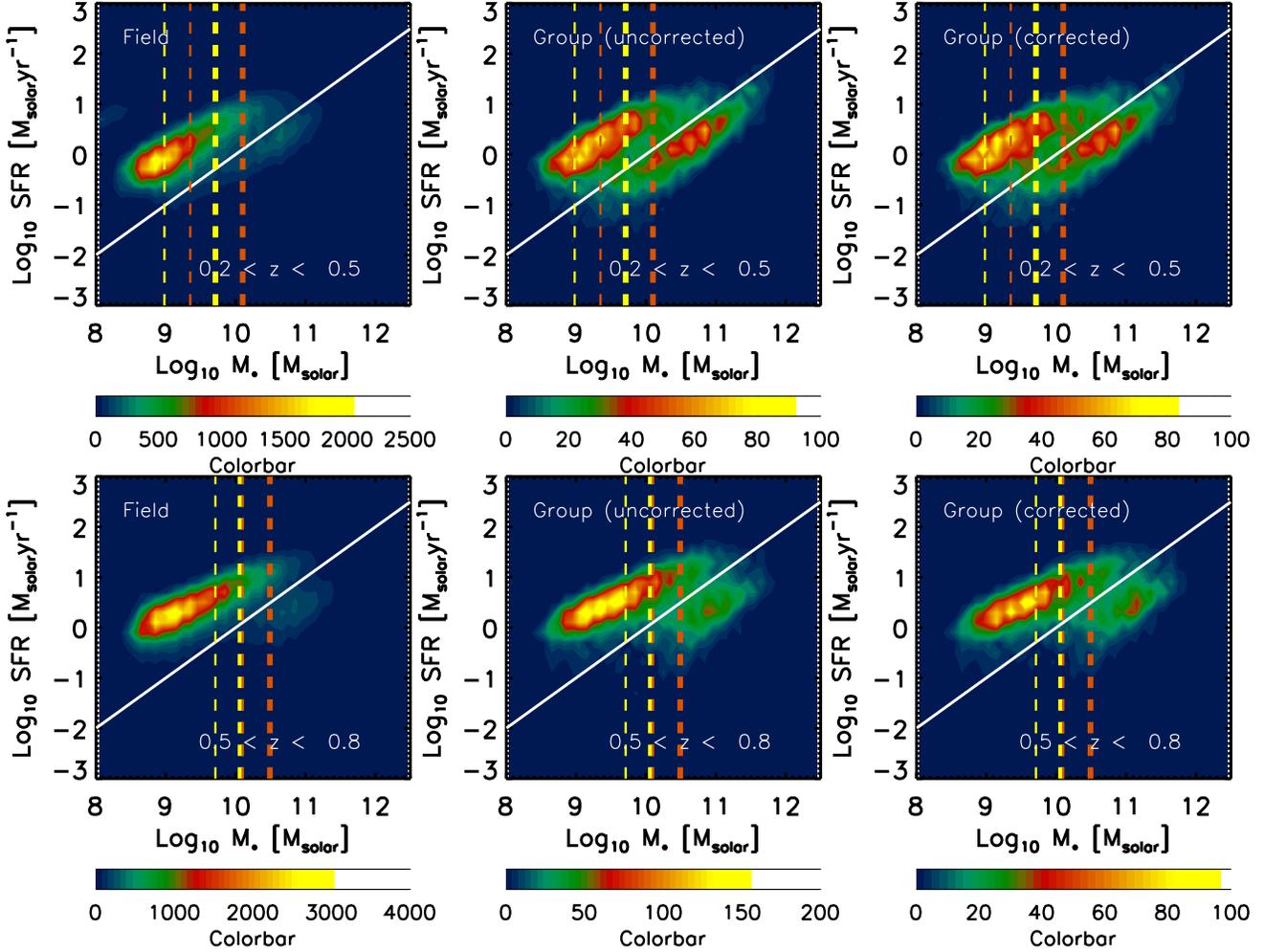}

\caption{SFR and \sm~ distribution of galaxies in two of the PS1MD fields (MD04 \& MD07), separated into bins of redshifts: $0.2 < z < 0.5$ (upper panels) and $0.5 < z < 0.8$ (lower panels). The colors are scaled according to the number of galaxies enclosed in each SFR and \sm~ grid. The left panels show results for field galaxies; the middle panels show results for galaxies identified in groups with masses in between $10^{13}$ and $10^{14}$ \Msolar~ by PFOF group-finding algorithm; the right panels give results for group galaxies after correcting for the completeness and contamination. The white solid line shows the dividing threshold (SSFR = $10^{-10}$yr$^{-1}$) separating the star-forming and quiescent population. 
The thick (thin) dashed lines represent the mass completeness limits for galaxies with the reddest colors in the  star-forming (yellow lines) and quiescent (red lines) populations at the upper- (lower-) redshift limits of each panel.   
\label{fig:sfrsm}}
\end{figure*}

\begin{figure*}[h]

\includegraphics[angle=-270,width=19cm]{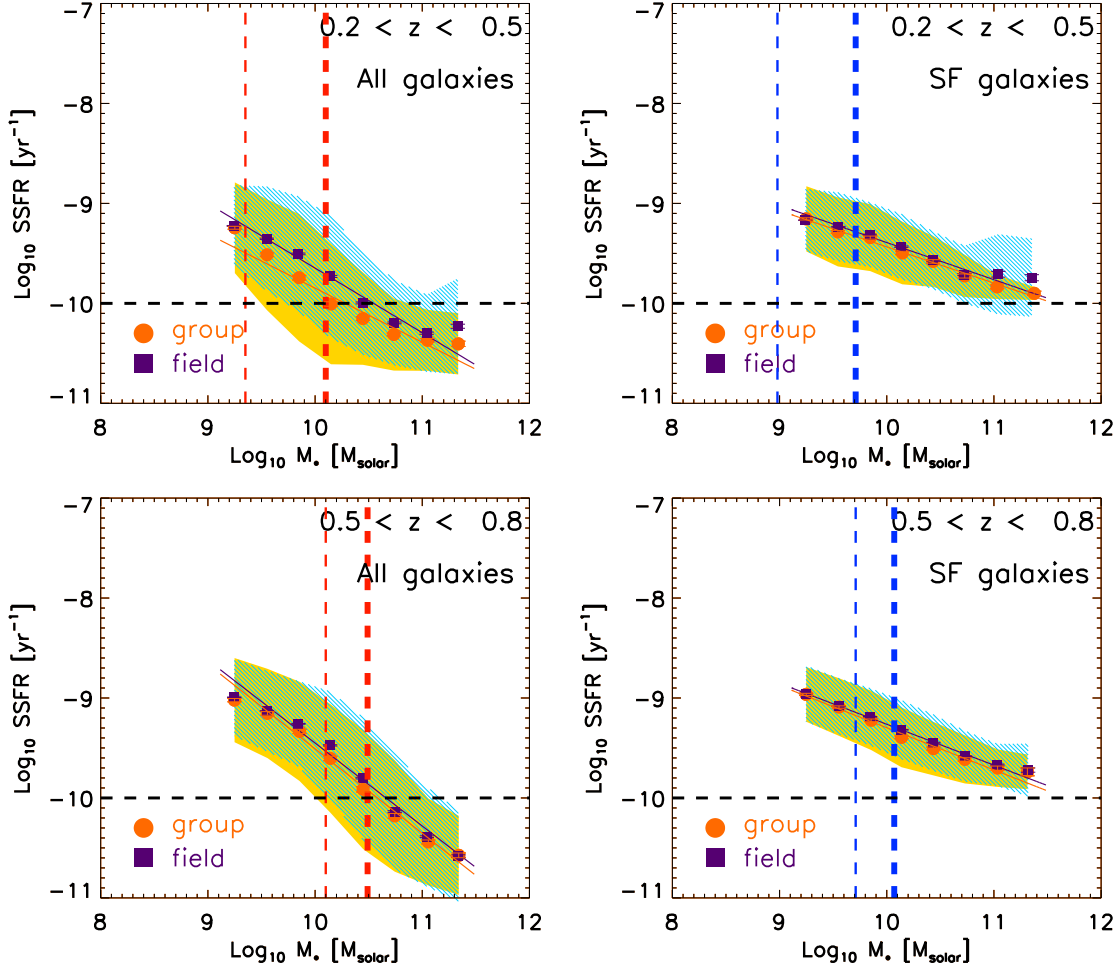}

\caption{SSFR--\sm~ relation for all galaxies (two left panels) and for star-forming galaxies only (two right panels) in two redshift bins. The black horizontal dashed lines represent the  SSFR threshold ($10^{-10}$ per year) that separates the 'star-forming' (SF) and 'quiescent' (non-SF) populations. Orange circles and purple squares show the means of the SSFR in each stellar mass bin for group and field galaxies respectively. The error bars denote standard errors (standard deviation divided by the square root of sample size in each bin). The shaded regions represent the 1-$\sigma$ standard deviation (yellow: groups; blue: field; green: the overlapped regions). The thick (thin) dashed lines represent the mass completeness limits for galaxies with the reddest colors in the  star-forming (blue lines) and quiescent (red lines) populations at the upper- (lower-) redshift limits of each panel.
\label{fig:ssfrsm}}
\end{figure*}

\begin{figure*}[h]

\includegraphics[angle=-270,width=19cm]{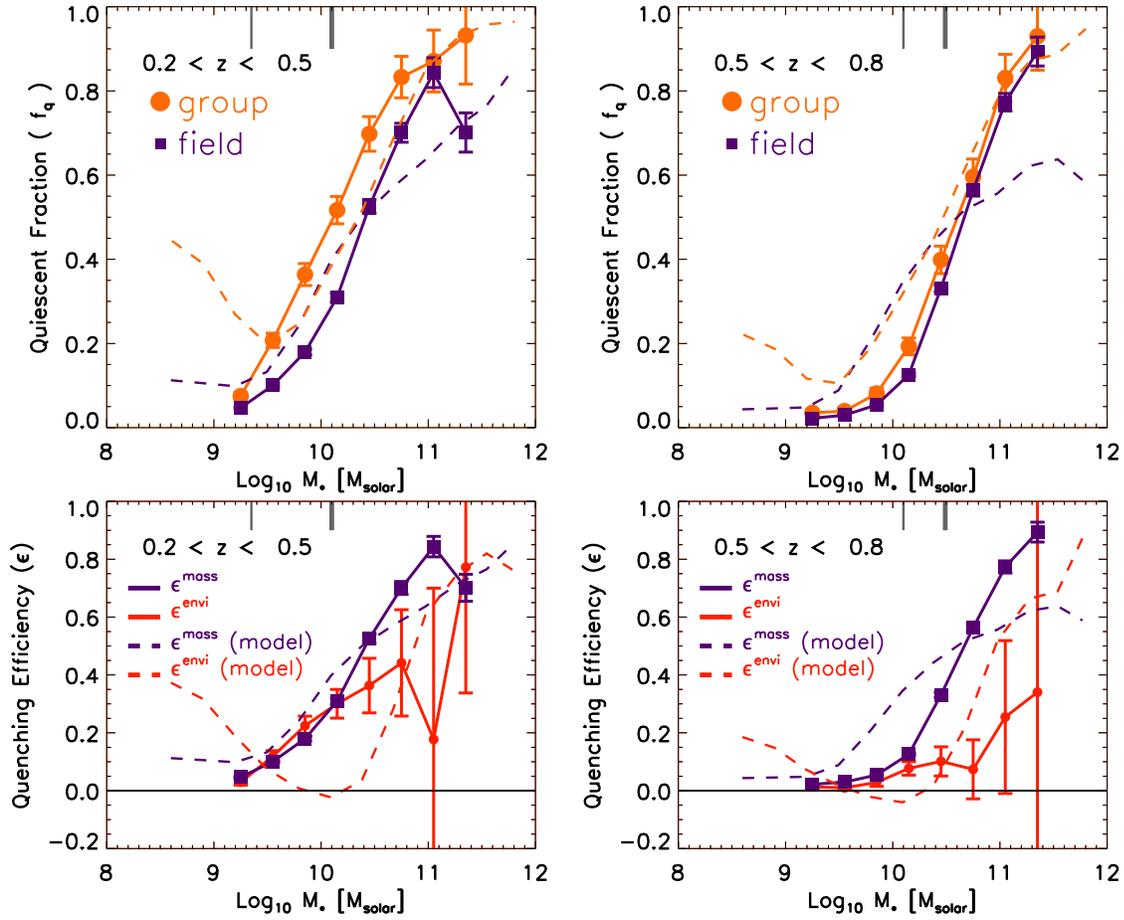}

\caption{Upper panels: the fraction of quiescent population as a function of stellar mass in groups (orange symbols) and in the field (purple symbols). The purple and orange dashed curves show the theoretical predictions using the \citet{fon08} model for galaxies located in the dark halos with masses of $10^{10}$M$_\odot$ $<$ M$_{halo}$ $<$ $10^{12.5}$M$_\odot$ and $10^{12.5}$M$_\odot$ $<$ M$_{halo}$ $<$ $10^{14}$M$_\odot$ respectively. Bottom panels: the environment quenching efficiency $\epsilon^{envi}$ (red curves) and mass quenching efficiency $\epsilon^{mass}$ (same as $f_{q}^{field}$; purple curves) as a function of stellar mass. The grey thick (thin) ticks shown in the upper side of each panel denote the mass completeness limits for the reddest colors of galaxies at the upper- (lower-) redshift limit of each panel. The purple and red dashed curves show the stellar mass quenching and environment quenching efficiency predicted in the \citet{fon08} model respectively. 
\label{fig:fq}}
\end{figure*}

\begin{figure*}[h]

\includegraphics[angle=-270,width=19cm]{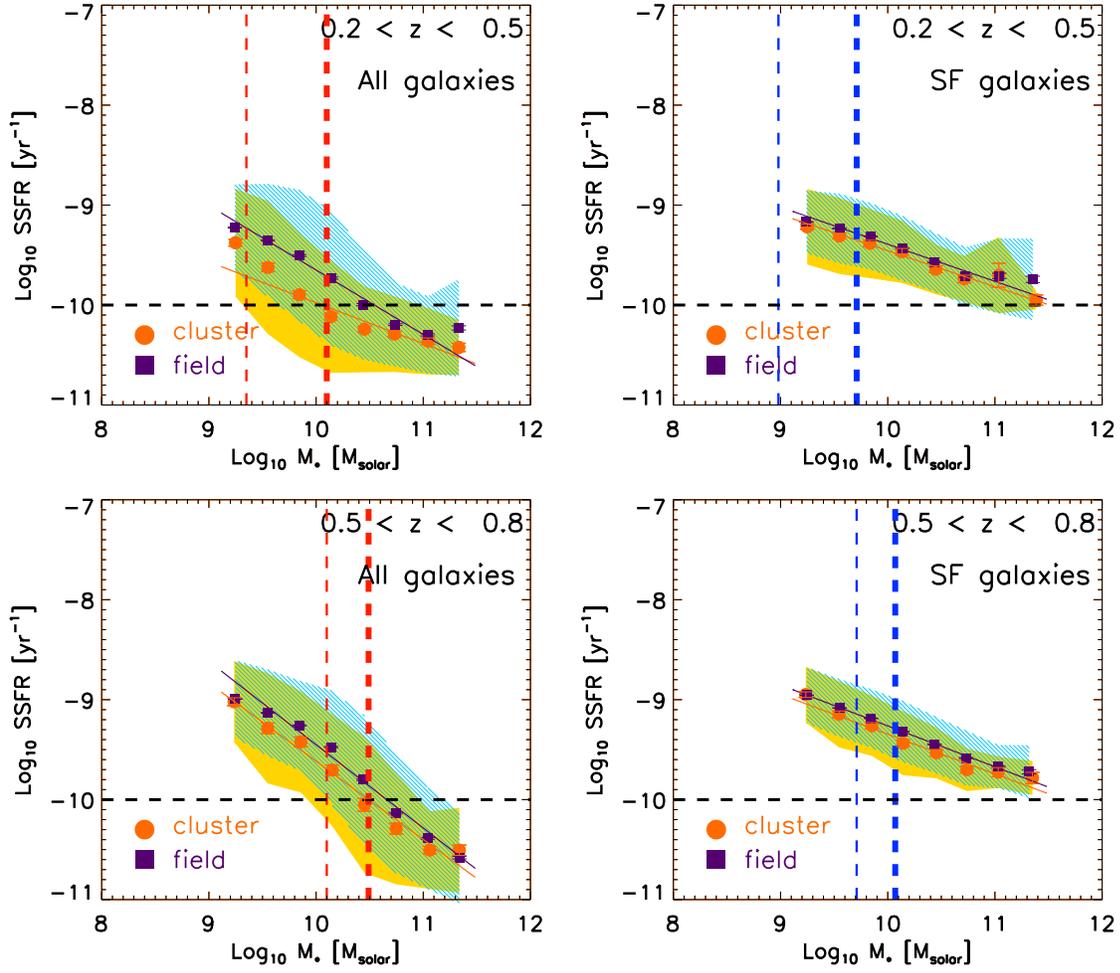}

\caption{Similar to Figure \ref{fig:ssfrsm} but for PFOF groups with $N_{rich} > 25$ (cluster-scale). 
\label{fig:ssfrsm_cluster}}
\end{figure*}

\begin{figure*}[h]

\includegraphics[angle=-270,width=19cm]{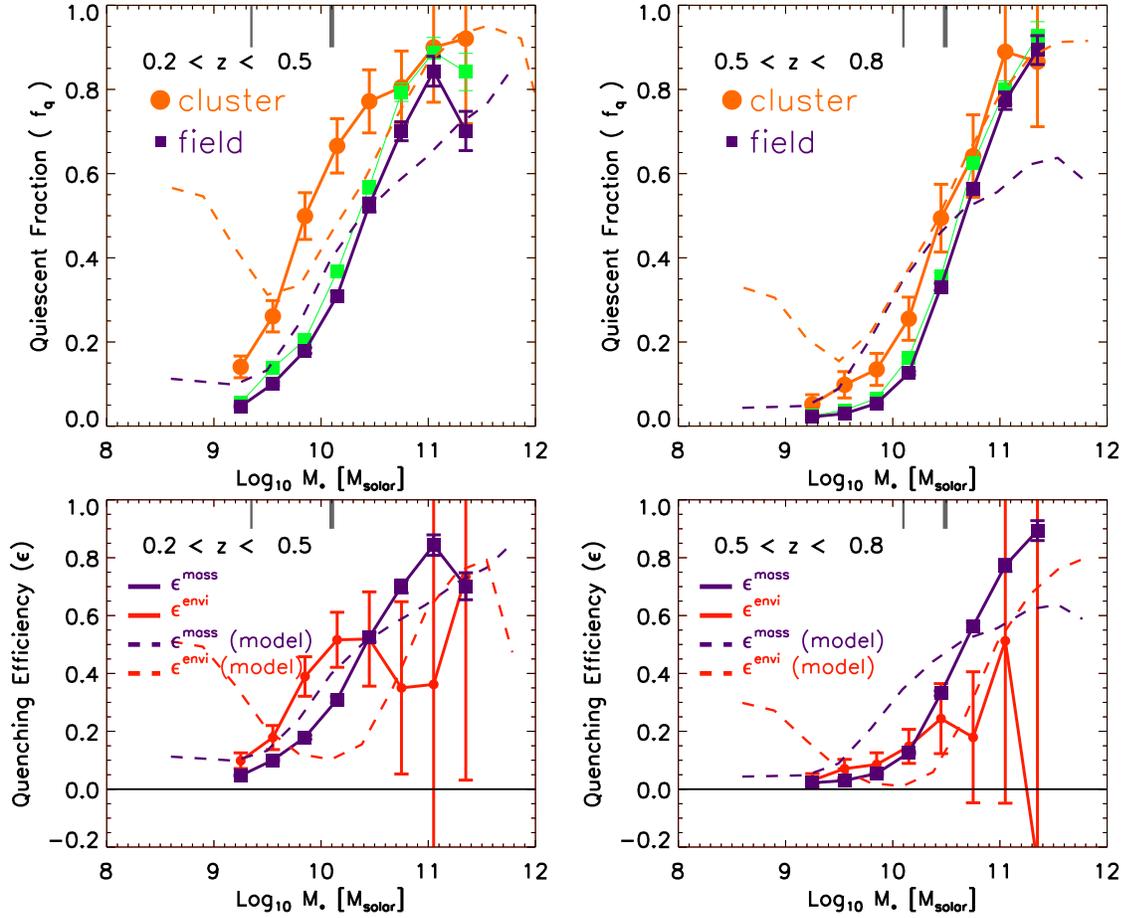}

\caption{Similar to Figure \ref{fig:fq} but for PFOF groups with $N_{rich} > 25$ (cluster-scale). The green data points present the quiescent fraction if the SFR of field galaxies is globally reduced by the amount of suppression seen in cluster SF galaxies (see text). The purple and orange curves show the theoretical predictions using the \citet{fon08} model for galaxies located in the dark halos with masses of $10^{10}$M$_\odot$ $<$ M$_{halo}$ $<$ $10^{12.5}$M$_\odot$ and $10^{14}$M$_\odot$ $<$ M$_{halo}$ $<$ $10^{16}$M$_\odot$ respectively.
\label{fig:fq_cluster}}
\end{figure*}

\begin{deluxetable}{lccc}
\tabletypesize{\scriptsize}
\tablewidth{0pt}
\tablecaption{Best-fitting parameters for the SFR--\sm~ relation of the star-forming sequence in the field and the groups.\label{tab:fit}}
\tablehead{
    \colhead{Subsample} &
    \colhead{Number$^{a}$} &
    \colhead{$\alpha$} &
    \colhead{$Log_{10}\beta$}
}

\startdata
Field   ($0.2 < z < 0.5$)           & \nodata & $0.629 \pm 0.007$ & $-5.682 \pm 0.068$ \\
Group   ($0.2 < z < 0.5$)           & 610 & $0.638 \pm 0.011$ & $-5.813 \pm 0.111$ \\
Cluster ($0.2 < z < 0.5$)           & 76 & $0.640 \pm 0.026$ & $-5.854 \pm 0.264$ \\
Field   ($0.5 < z < 0.8$)           & \nodata & $0.591 \pm 0.003$ & $-5.170 \pm 0.034$ \\
Group   ($0.5 < z < 0.8$)           & 875 & $0.578 \pm 0.012$ & $-5.080 \pm 0.119$ \\
Cluster ($0.5 < z < 0.8$)           & 61 & $0.602 \pm 0.026$ & $-5.370 \pm 0.264$ 
\enddata

\tablecomments{$^{(a)}$: This column denotes the number of groups and clusters used in the analysis of each subsample.}

\end{deluxetable}

\section{DATA, SAMPLE SELECTIONS, AND METHODS \label{sec:data}}
\subsection{Pan-STARRS data}
Pan-STARRS 1 (hereafter PS1) is a 1.8 meter telescopes equipped with a CCD digital camera with 1.4 billion pixels and 3-degree field of view, located on the summit of Haleakala on Maui in the Hawaii Islands \citep{ona08,kai10}. The \PS\ observations are obtained through a set of five broadband filters, which we have designated as \gps, \rps, \ips, \zps, and
\yps. Under certain circumstances \PS\ observations are obtained with
a sixth, ``wide'' filter designated as \wps\ that essentially spans
\gps, \rps, and \ips. Although the filter system for \PS\ has much in
common with that used in previous surveys, such as SDSS \citep{SDSS}, there
are important differences. The \gps\ filter extends 20~nm redward of
$g_{SDSS}$, paying the price of 5577\AA\ sky emission for greater
sensitivity and lower systematics for photometric redshifts, and the
\zps\ filter is cut off at 930~nm, giving it a different response than
the detector response defined $z_{SDSS}$.  SDSS has no corresponding
\yps\ filter.  Further information on the passband shapes is described
in \cite{stu10}. 

There are two major components of the PS1 survey which started observations in 2010: the 3$\pi$ survey and the Medium Deep Survey (MDS). The largest portion of time is devoted to the 3$\pi$ survey (Chambers et al., in preparation), which is scanning the entire sky north of declination $-30\deg$ in five filters, \grizy \ \citep{ton12b}, in six separate epochs spanning $\sim$3.5 yr, each epoch consisting of a pair of exposures taken $\sim$25 min apart. Each field center is visited by a total of 20 exposures per year in all filters. By
stacking all these exposures, PS1 will provide a 30 000 deg$^{2}$ survey
of the sky to a depth expected to be somewhat greater than that of
the Sloan Digital Sky Survey (SDSS; York et al. 2000), especially at
redder wavelengths. The survey is expected to be completed early
in 2014 and publicly released to the world in 2015. More details on the characteristics of the 3$\pi$ are described in \citet{mag13} and \citet{met13}. 

This paper uses images and photometry from the Pan-STARRS1 Medium-Deep Field survey. The PS1 Medium Deep Fields (MD fields) consist of 10 spatially well-separated fields, each with 7 square degrees. MD fields are observed repeatedly with $g_{\mathrm{P}1}r_{\mathrm{P}1}i_{\mathrm{P}1}z_{\mathrm{P}1}y_{\mathrm{P}1}$ filters \citep{stu10,ton12a}, with the goal to reach \{26.0, 26.0, 26.3, 25.6, 24.3\} in AB magnitude respectively for point sources after the 3.5-year period of the PS1 mission finishes.  Observations of the Medium-Deep fields are taken each night,
cycling through the various Pan-STARRS1 filters, during that
portion of the year that the fields are accessible at less than
1.3 airmasses. A nightly observation in a given filter consists
of eight dithered exposures, with a typical cadence as shown
in Table 3 of \citet{ton12a}.

Nightly stacks of PS1 data are produced by Image Processing Pipeline \citep[IPP;][]{mag06}. The Pan-STARRS1 IPP system performed flatfielding on each of the individual images, using white light flatfield images
from a dome screen, in combination with an illumination correction obtained by rastering sources across the 
field of view. Bad pixel masks were applied, and carried forward for use in the stacking stage. After determining 
an initial astrometric solution, the flat-fielded images were then warped onto the tangent plane of the sky, using a flux
conserving algorithm. The plate scale for the warped images is 0.200 arcsec/pixel. 

Using the ``Astromatic'' software \footnote{http://www.astromatic.net}, SCAMP \citep{ber06}, SWarp \citep{ber02} , and SExtractor \citep{ber96}), we produced our own version of deep stacks and associated catalogs based on all the nightly stacks generated by IPP between May 2010 and December 2011. The zeropoint of the photometry is calibrated against the SDSS-DR7 catalog. In addition, all PS1 MD fields are covered in CFHT MEGACAM $u^{*}$-band taken by Eugene Magnier et al. as part of the PS1 efforts. We have also downloaded the calibrated  $u^{*}$-band images from the CADC Archive system and produced deep stacks and catalogs following a similar process as for the PS1 images. The final six-band master catalogs based on the $i_{\mathrm{P}1}$-band detected objects are generated by running SExtractor in dual mode. Because the median seeing varies from 0.8$\arcsec$ to 1.1$\arcsec$ across different bands, the fluxes measured using a fixed size of aperture sample different fractions of lights of galaxies. Therefore we have chosen to use the $AUTO$ magnitude of the $i_{\mathrm{P}1}$ band to be the total $i_{\mathrm{P}1}$-band magnitude. The colors are defined as the difference in the $ISO$ magnitude that are output from SExtractor using the isoarea defined by the $i_{\mathrm{P}1}$ band. Empirically we find that this approach yields the best performance of photometric redshift.  This work makes use of data taken in two of the PS1 MD fields, namely MD04 and MD07, which cover well-known multi-wavelength extra-galactic fields, COSMOS and Extended Groth Strip respectively. The choice of these two fields is primarily driven by the availability of large number of redshifts, robust redshift identification, and high sampling rate of the zCOSMOS \citep{lil07} and DEEP2  \citep{new13} spectroscopic redshift samples which are crucial for the purpose of calibrating our photometric redshifts (see \S \ref{sec:photoz}) and our group-finder algorithm (see \S \ref{sec:PFOF}). In the current version of the MD07 catalog used in this work, we reach the depth of \{25.63, 25.05, 24.95, 25.03, 24.46, 23.18\} for \{$u^{*}g_{\mathrm{P}1}r_{\mathrm{P}1}i_{\mathrm{P}1}z_{\mathrm{P}1}y_{\mathrm{P}1}$\} at 5$\sigma$ using the 1$\arcsec$ aperture in radius. The depth of MD04 is comparable to MD07, except for $y_{\mathrm{P}1}$ which is $\sim$ one  mag shallower. Details are given in a companion paper by S. Foucaud et al. (2013, in prep.). 

\subsection{Photometric redshifts}\label{sec:photoz}
Photometric redshifts (hereafter \photoz, or sometimes $z_{ph}$) are computed by fitting the six optical bands including PS1 $grizy$-band  and CFHT $u^{*}$ photometry with the publicly available EAZY code \footnote{http://www.astro.yale.edu/eazy/}\citep{bra08}. The templates adopted in this work are taken from one of the template sets provided by another public \photoz~ software ``LePhare'' \citep{arn99,ilb06}\footnote{http://www.cfht.hawaii.edu/~arnouts/lephare.html}, called 'CFHTLS-SED', which includes 66 templates originally constructed by \citet{ilb06} based on four observed galaxy spectra from \citet{col80} and two starburst galaxy spectra from \citet{kin96} to optimize the photometric redshift results for the CFHTLS dataset. The details are described in \citet{ilb06} and \citet{cou09}. 

The computation of \photoz~ involves several steps. First we ran EAZY only for galaxies with secured spectroscopic redshifts to determine the systematic zeropoint offsets in each band by measuring the medians of the differences in the photometry between the data and the best-fit templates. Then we applied the derived zeropoint offsets back to the data and measured the \photoz~ for all galaxies. The derived systematic zerpoint offsets are often negligible in $u^{*}$ (-0.01 mag), $g_{\mathrm{P}1}$ (0.00 mag), $r_{\mathrm{P}1}$ (-0.01 mag), and $z_{\mathrm{P}1}$ bands (0.00 mag), and are slightly larger in $i_{\mathrm{P}1}$ (-0.05 mag) and $y_{\mathrm{P}1}$ (-0.059 mag) bands. In order to reduce the catastrophic outliers due to the confusion between the Lyman and Balmer breaks in the absence of near-infrared data, we have adopted a prior on the redshift distribution for any given range of $i$-band magnitude using a mock galaxy catalog constructed based on a semi-analytical model described in \citet{guo10}. 

Two large spectroscopic redshift surveys, the DEEP2 \citep{new13} and zCOSMOS \citep{lil07} samples, are used to calibrate the zeropoints and to characterize our \photoz~ performances in MD07 and MD04 respectively. Figure \ref{fig:photoz} compares our \photoz~ results against the spectroscopic redshifts in these two MD fields. Following the definition adopted by \citet{ilb06}, we quantify the \photoz~accuracy (hereafter \deltaz) using the normalized median absolute deviation \citep[NMAD][]{hoa83}, defined as 1.48$\times$median($\Delta~z$/($1+z_{s}$)), where $z_{ph}$ is the \photoz, $z_{s}$ is the spectroscopic redshifts, and $\Delta z = z_{ph} - z_{s}$. The outlier rate (hereafter \outlier) is defined as the fraction of galaxies with $\Delta~z > 0.15\times(1+z_{s})$. $\sigma$ and \outlier~ are found to be 0.047 and 4\% at $i < 22.5$ in MD04, and 0.051 and 7\% at $r < 24.1$ in MD07.

A complementary way of assessing the performance of the PS1 $z_{ph}$ is to use overlapping photometric catalogues using a larger number of filters, thus providing more accurate values of $z_{ph}$. 
In this way, we can reproduce the same selection as in the PS1 catalogues, avoiding the possible biases introduced by the use of spectroscopic redshifts.
We used the COSMOS v1.7 catalogue of Ilbert et al. (2009, 2010), which covers $2 \deg^2$ in field MD04, and contains photometric redshifts ($z_{ph}^{cosmos}$) computed with LePhare using 31 broad- and narrow-band filters.
The error in the $z_{ph}^{cosmos}$ ($\sigma \leq 0.011$ for $i^{+} < 24$ and $z < 1.5$) is much smaller than that expected in PS1, so it is a good reference point for our comparison.
For the limit used in this work, $i_{\mathrm{P}1} < 24$, we found $93733$ objects which were present in the COSMOS catalogue. 
We define the dispersion ($\sigma^p$) and outlier rate ($\eta^p$) of the $z_{phot}$ in this case in the same way as above, now taking $\Delta z = z_{ph} - z_{ph}^{cosmos}$, and obtain $\sigma^p = 0.083$ and $\eta^p = 20.8\%$ for the full catalogue.
We also define the bias in the estimation of $z_{ph}$ as the median of $\Delta z /(1 + z_{ph}^{cosmos})$, obtaining a value of $bias = 0.014$.
We show the comparison between the PS1 and COSMOS redshifts in Fig.~\ref{fig:photoz}.
When restricting the comparison to objects with $i_{\mathrm{P}1} < 22.5$ we found $\sigma^p = 0.050$, $\eta^p = 4.5\%$, $bias = 0.012$.
We can use this comparison to characterize the redshift bins used in this work for galaxies with $i_{\mathrm{P}1} < 24$. To this end, we introduce the bin outlier rate $\eta_{bin}$, defined as the fraction of objects selected in a given bin with a value of $z_{ph}^{cosmos}$ farther than $1 \sigma^p$ from the bin limits.
For the first bin, $0.2 < z_{ph} < 0.5$, we found $\sigma^p = 0.065$, $bias = 0.027$ and $\eta_{bin} = 9.0\%$.
For the bin in $0.5 < z_{ph} < 0.8$ we obtain $\sigma^p = 0.079$, $bias = 0.022$ and $\eta_{bin} = 23.5\%$.

As a further test, we made a comparison with the `Gold' catalogue of the ALHAMBRA survey (Molino et al., 2013), which provides $z_{phot}$ down to $I < 23$ obtained using 23 bands (with an error $\sigma = 0.010$), and overlaps an area of $\sim 0.25 \deg^2$ in MD04 and $\sim 0.5 \deg^2$ in MD07.
We found, for $i_{\mathrm{P}1} < 22.5$, the values $\sigma^p = 0.048$, $\eta^p = 5.3\%$ and $\sigma^p = 0.047$, $\eta^p = 5.1\%$ for MD04 and MD07, respectively.
We therefore conclude that the quality of our $z_{phot}$ is consistent in the two fields considered. We restrict this study to the two MD catalogs (MD04 and MD07) which consist of 313,997 galaxy-like objects brighter than $i_{\mathrm{P}1}$ = 24 mag over 14 square degrees in the redshift between 0.2 and 0.8.

\subsection{K-correction, stellar mass and star formation rate \label{sec:kcorr}}
The K-correction which converts the observed magnitudes to the restframe magnitudes is computed following a similar approach as described in \citet{wil06}. At a given redshift, we fit a polynomial to the relationship between the K-correction term (using the observed bandpass closest to the desired restframe quantity) and a pair of adjacent observed color based on empirical templates taken from \citet{kin96}. Depending on the redshift, the corresponding polynomial formula is then applied to each galaxy in the redshift range of $0 < z < 1.45$. 

We derive stellar masses by fitting the broad-band photometry to the
synthesized templates generated with \citet{bru03} models using the
SED fitting code 'FAST' \citep{kri09}. During the fitting, we fix the
metallicity to be solar value \footnote{To test the effect of metallicity, we compared the stellar masses derived with and without fixing the metallcity to be the solar value for a subsample with known spectroscopic redshifts, and found that the scatter between these two cases is $\sim 0.1$ dex.} and the redshift to be the \photoz~
determined from EAZY, while the rest of model properties including the
age, star formation time scale $\tau$ (assuming an exponentially decaying star formation history), and dust content $A_{V}$ are treated as free parameters. 

For a subset of galaxies overlapped with the COSMOS field, we compared our derived stellar masses based on SED fitting to the 6-band PS1 photometry to the ones in the COSMOS v1.7 photometric redshift catalog \citep{ilb10}, computed with LePhare using the 31 broad- plus narrow-band COSMOS photometry data based on the same \citet{bru03} models. The COSMOS stellar masses were previously derived using the Chabier IMF, and hence we have multiplied those by a constant factor of 1.8 to covert them in to Salpeter IMF. In Fig. \ref{fig:sm} we show the comparison for galaxies (after correcting for the difference in the adopted IMF) whose PS1 $z_{ph}$ are consistent with COSMOS $z_{ph}$ within the typical PS1 \photoz~ uncertainty (i.e., ($z_{ph}^{ps1} - z_{ph}^{cosmos}$)/(1 + $z_{ph}^{cosmos}$) < 0.05). We found that the scatter between the two measurements for ``good'' photometric redshift sample is $\sim0.23$ dex.

To estimate the stellar mass completeness of our sample, we first translate the 5-$\sigma$ limiting magnitudes in the observed PS1 bands into the rest-frame quantities for galaxies at a given redshift, and then estimate the corresponding stellar mass using the empirical formula obtained by \citet{lin07} that relates the rest-frame magnitudes and colors to the stellar mass. For a fixed rest-frame magnitudes, the stellar mass is greater for galaxies with redder colors. Therefore we take the reddest colors of star-forming and quiescent populations respectively when computing the mass limit of the above two samples.  This yields a mass limit log$_{10}$(\sm/\Msolar) = 9.4 (9.0), 10.1 (9.7), and 10.5 (10.1) for red (blue) galaxies at $z \sim 0.2$, $z \sim 0.5$, and $z \sim 0.8$ respectively. The main sample used in this study contains 244,338 galaxies which have their stellar masses greater than $10^{9}$\Msolar.

Unlike stellar masses, star formation rates are rather sensitive to
the degeneracies of the parameters in the SED fitting procedure in the
case when there is not enough longer wavelength data. Instead of using
the SFR output from FAST, we derived the SFR by adopting the approach described in \citet{mos12} which parameterizes the SFR as a function of rest-frame optical U and B magnitudes (see Eq. (1) and Table 3 in their paper) by calibrating against the [OII] emission line luminosities in the DEEP2 redshift sample \citep{new13}. One caveat of this method is that it may be biased against highly dust-obscured populations \citep{mos12}. However, as illustrated in \citet{mos12}, it uncovers the SFR of galaxies with a wide range of star formation activities, including both blue and red galaxies and therefore is suitable for our purposes in a statistically representative way.

\subsection{Group identification}\label{sec:PFOF}
Proper identification of galaxy groups is essential to achieve our
goals of studying galaxy properties in different environments. While
spectroscopic redshift surveys have adequate redshift resolution to
secure the group members, in practice it is observationally expensive
to conduct a large-volume spectroscopic redshift survey with high
sampling rate to yield statistically meaningful large numbers of
groups of galaxies. Moreover, the spectroscopic sample has the
tendency to be biased toward emission-line galaxies due to the greater
S/N in the line measurement for the redshift identification, which may
possibly bias the results in particular for faint galaxies. In
contrast, multi-band imaging surveys are relatively efficient at
obtaining a large size of galaxy sample. However, group/cluster
identification is challenging due to the lack of the redshift
precision. One commonly adopted cluster identification method is the
so-called `red-sequence' method which has proven to be successful in
finding galaxy clusters by  pin point the cluster redshifts \citep{gla05}. Nevertheless, this method relies on identifying the tight sequence in the red population and hence breaks down for group scales and suffers difficulty in recovering blue group members. 

An alternative method, Probability Friend-of-Friend group finder
(hereafter PFOF; \citet{liu08}), attempts to recover both the blue and
red populations in groups. PFOF considers the probability distribution
function of photometric redshifts of each galaxy and computes the
likelihood of a given pair of galaxies being spatially associated.
The group sample is then constructed by linking galaxy pairs that have the likelihood exceeding a certain threshold, which is a free parameter determined in the process of optimizing the group memberships to the known spectroscopically-identified groups and clusters in the same field. 

The PFOF group finder has been tested intensively using galaxy mock catalogs as well as observational datasets \citep{liu08,jia13}. In \citet{jia13}, we applied PFOF to the MD04 and MD07 photometric redshift catalogs, calibrated using the zCOSMOS and DEEP2 spectroscopically-identified galaxy groups \citep{kno09,ger12} respectively. The details are described in \citet{jia13}. In this work, we make use of an updated version of the PFOF-generated group samples in MD04 and MD07 with the difference that instead of using different subset training samples in different fields, the two PFOF parameters, linking lengths in both the projected sky plane and the line-of-sight direction, and the probability threshold are trained by optimizing the purity and completeness against the DEEP2 spectroscopic identified group sample in the MD07. The same set of parameters are applied to both MD04 and MD07 as the photometric redshift uncertainty is comparable between the two fields. 

Fig. \ref{fig:grouphist} shows the histogram of the richness ($N_{rich}$) for the 68,180 PFOF groups with $N_{rich} > 2$ identified in MD04 and MD07 in the redshift range of $0.2 < z < 0.8$, where $N_{rich}$ is defined as the total number of members brighter than $i = 24$ in a group. In this work, we divide our PFOF group samples into two subsets, one with $10 < N_{rich} < 25$ (the 'group' sample) and the other with $N_{rich} > 25$ (the 'cluster' sample). For the group sample, the richness cut roughly translates into a group mass of $10^{13.2} < M_{halo} < 10^{13.8}$ \Msolar~ at $z \sim 0.4$ and $10^{13.4} < M_{halo} < 10^{14.0}$ \Msolar at $z \sim 0.8$ respectively. We consider field galaxies as those not associated with any groups with $N_{rich}\ge 2$. For any given group-finding method, it is important to characterize its capability of recovering true group memberships as well as the contamination rate from the field galaxies. \citet{jia13} has shown that despite the \photoz~ uncertainty is in general worse for blue galaxies compared to that for red galaxies, with proper tuning of the linking length and the likelihood threshold, PFOF is able to recover blue members at a similar level as for red members. On the other hand, the contamination from blue galaxies in the field is typically larger than that from red galaxies. The reason for that is because the linking length can not be too small otherwise we lose many blue members. Therefore the PFOF parameter optimization often leads to a linking length which is large enough to recover both blue and red galaxies at a similar level. The trade off is that the contamination rate from blue galaxies would become larger accordingly. This selection effect needs to be corrected when comparing the galaxy properties in different environments.

\section{RESULTS}
\subsection{SFR $-$ \sm~ relation of galaxies in fields versus groups \label{sec:sfrsm}}

Fig. \ref{fig:sfrsm} shows the SFR -- \sm~ distribution for our sample
of galaxies in the PS1 MD04 and MD07 fields, separated into bins of redshift: $0.2 < z < 0.5$ (upper panels) and $0.5 < z < 0.8$ (lower panels). The colors are scaled according to the numbers of galaxies enclosed in each SFR and \sm~ grid. The galaxies are further divided into `field' and `group' populations according to the PFOF identification (left-hand and middle panels). The stellar mass limits corresponding to galaxies at the upper- and lower- redshift limits, shown as vertical dashed lines, are computed using the method described in \S\ref{sec:kcorr}.

As mentioned in \S\ref{sec:PFOF}, the PFOF group catalogs are neither complete nor pure in terms of the memberships which may potentially bias our results when comparing the SFR $-$ \sm~ distribution of galaxies in different environments. To characterize the selection function for star-forming and quiescent group members, we compute the recovery rate $R_{r}$ and contamination rate $R_{c}$ for star-forming and quiescent populations separately in bins of redshift and stellar mass, by cross-referencing the MD07 PFOF group catalogs to the spectroscopically-identified group catalogs constructed in the EGS field (part of MD07) by \citet{ger12}. $R_{r}$ is calculated as the fraction of spectroscopically-identified group members that are also PFOF members, while $R_{c}$ is computed as the fraction of PFOF group members that are not associated with  spectroscopically-identified groups. We provide more details in Appendix \ref{app:corr}.

For each of the SFR and \sm~ grid, we then correct for both the incompleteness and contamination effects by subtracting the field contribution as follows:

\begin{equation}\label{eq:sfrsm}
n^{g}(SFR,M_{*}) = \frac{1 - R_{c}}{R_{r}}\times n_{raw}^{g}(SFR,M_{*}),
\end{equation}
where $n^{g}(SFR,M_{*})$ denotes the numbers of group galaxies for a given SFR and \sm~ grid. Ideally $R_r$ and $R_c$ should be computed in a SFR and \sm~ grid size as small as possible. However, our calibration relies on the limited numbers of spectroscopic group members, so we have chosen to compute $R_r$ and $R_c$ in six regions in the SFR and \sm~ space, and then for each region we apply the same global value to each SFR and \sm~ grid. The `corrected' group populations are shown in the right-hand panels of Fig. \ref{fig:sfrsm}.

In Fig. \ref{fig:sfrsm}, it can be seen that the distributions of field and group galaxies in the SFR and \sm~ plane are distinct. Qualitatively, the stellar masses of group galaxies are systematically shifted toward higher masses for both the star-forming and quiescent populations. In addition, the presence of the quiescent sequence is more prominent in the group environments, especially in the lower-redshift bin. This is in general in good agreement with previous studies concluding that the red fraction of galaxies is greater in the group than in the field \citep{ger07,bal09,gio12}. For further analysis used in this work, we define star-forming and quiescent galaxies as those with SSFR greater and lower than $10^{-10}$ per year respectively.

It is worth noting that in the high-redshift bin, there exists a class of massive star-forming galaxies with stellar mass $> 10^{10}$ \Msolar~ in the group which nevertheless is rare in the field environments in the similar redshift range. The presence of massive blue galaxies (or equivalently massive galaxies with high SFR) in dense environments at $z \sim 1$ has also been constantly reported in other studies \citep{ger07,coo07,tra10,koy13}, suggesting a more advanced mass assembly stage in high-density regions. Our results suggest that while the fraction of quiescent galaxies in group environments is larger, galaxy groups, on average, are a preferential environment for the formation of massive star-forming galaxies.

To quantitatively describe the SFR $-$ \sm~ relation of the
main-sequence in our sample, we fit the data for the star-forming
galaxies with a linear relation between $\log_{10}\,$SFR and
$\log_{10}\,$\sm: $\log_{10}{\rm SFR} =  \alpha \times\log_{10}$ \sm + $\log_{10}\beta$, where the slope $\alpha$ and the amplitude $\beta$ are both fitting parameters. The best-fit results are given in Table \ref{tab:fit}. The best-fit slopes of the main sequence (star-forming population) in our field sample are between 0.59 and 0.63, in broad agreement with previous results at similar redshifts \citep{noe07,whi12}, but slightly shallower compared to those found at $z \sim 2$ \citep{dun09,pan09,lin12}. The amplitude of the SFR $-$ \sm~ relation in the lower-z bin is also in good agreement with the results from \citet{noe07} who used the AEGIS sample at $0.2 < z < 0.7$, after correcting for the difference in the IMF. From Table \ref{tab:fit}, we also see that both the slope and the amplitude between the field and group `main-sequence' are nearly the same, and the difference is only seen when the quenched population is included. We will discuss this further in the next section.

\subsection{The SSFR $-$ \sm~ relation of star-forming galaxies \label{sec:ssfrsm}}

To have a closer view of the stellar mass dependence of the SFR, we measure the mean of SFR normalized by the stellar mass, namely, specific star formation rate (SSFR), in each stellar mass bin. These measurements are listed in Table \ref{tab:ssfrfq}. Galaxies are divided into the `star-forming' (SF) and `quiescent' (non-SF) populations by their specific star formation rate with a threshold of $10^{-10}$ per year. Fig. \ref{fig:ssfrsm} shows the SSFR as a function of \sm~ for all galaxies (left panels) and for star-forming galaxies (right panels), separated according to their environments (field: purple symbols; groups: orange symbols). The error bars denote the standard errors (standard deviation divided by the square root of sample size in a given stellar mass bin), while the shaded areas show the dispersion of the SSFR at a given stellar mass. The slope of the SSFR $-$ \sm~ relation becomes steeper when quiescent galaxies are included. While the difference in the median SSFR between field galaxies and group galaxies is apparent for the whole populations (left panels of Fig. \ref{fig:ssfrsm}), it becomes more negligible when the quiescent galaxies are excluded (right panels of Fig. \ref{fig:ssfrsm}). In other words, the group environment has little effect on the averaged star formation activities for `main-sequence' galaxies, whereas its effect is primarily on moving the galaxies out of the star-forming sequence toward the quiescent populations, leading to a suppressed of the mean SSFR of all (SF plus quiescent) group galaxies. 

The difference in the SSFR $-$ \sm~relation for `all' galaxies between field and groups, however, becomes smaller with increasing redshift. This is consistent with the trend found in previous works that suggest a convergence of galaxy properties between field and groups at $z \sim 1$ by studying the colors and the red fractions of group galaxies as a function of redshift \citep{ger07}.

\subsection{The stellar mass and environment dependence of quiescent fraction and quenching efficiency \label{sec:qf}}

In the previous section, we have shown that for a given \sm~ the mean SSFR of group galaxies is lower compared to field galaxies only if quiescent galaxies are included in the analysis. We next turn to discuss how the fraction of quiescent galaxies depends on the stellar mass and environment (Table \ref{tab:ssfrfq}). Fig. \ref{fig:fq} displays the quiescent fraction $f_{q}$ as a function of stellar mass in the field (purple symbols) versus groups (orange symbols). In both the redshift bins ($0.2<z<0.5$ and $0.5<z<0.8$), it can be seen that \fq~ increases rapidly with the stellar mass and this trend is independent of environment, consistent with previous works \citep{qua12,muz12,kov13}. In addition, \fq~ is in general higher in groups than in the field at fixed stellar mass. 

To quantify the excess of quenching due to pure environment effects, it is useful to compute the so-called ``environment quenching efficiency'' by removing the stellar mass dependence. Assuming that the ``mass quenching'' and ``environment quenching'' are independent, the fraction of quenched galaxies in the group environments can be expresses as follows:
\begin{equation}\label{eq:bluefra}
f_{q}^{group} = 1 - (1 - \epsilon^{envi})*(1 - \epsilon^{mass}),
\end{equation}
where $\epsilon^{envi}$ is the environment quenching efficiency, and $\epsilon^{mass}$ is the mass quenching efficiency. Assuming that only the mass quenching is in effect in the field environment, $\epsilon^{mass}$ is equivalent to $f_{q}^{field}$. Therefore, one can rewrite the environment quenching efficiency $\epsilon^{envi}$ as:
\begin{equation}\label{eq:epsilon}
\epsilon^{envi} = (f_{q}^{group} - f_{q}^{field})/(1 - f_{q}^{field}).
\end{equation}
As a result, $\epsilon^{envi}$ is identical to the fraction of galaxies that would be star-forming but which are however quenched in high-density regions as defined in \citet{pen10,qua12}.

Previous works have suggested that $\epsilon^{envi}$ has little dependence on stellar mass out to $z \sim 2$ \citep{pen10,qua12}, meaning that the environment quenching acts at a similar level regardless of the stellar masses of galaxies. Nevertheless we note that the sample sizes used beyond the local Universe were still very small. In the bottom panel of Fig. \ref{fig:fq}, we plot $\epsilon^{envi}$ (red curves) as a function of stellar mass for the PS1MD sample (also see Table \ref{tab:qe}). In the redshift range we are probing, we find a trend that $\epsilon^{envi}$ slightly increases with increasing stellar mass, different from the weak (or no) stellar mass dependence found in the zCOSMOS sample \citep{pen10} and in the UKIDSS UDS sample \citep{qua12}. We also find that the level of $\epsilon^{envi}$ becomes weaker in the higher-redshift bin, suggesting that the act of environment quenching operates more strongly in local Universe than at higher redshifts at a fixed stellar mass.

\subsection{Star-forming sequence in clusters \label{sec:cluster}}
While the main focus of this work is to study the impact of the group environment on galaxies, we are also able to conduct a similar analysis for clusters that are found in MD04 and MD07 (see Table \ref{tab:ssfrfq}), thanks to the large cosmic volume probed by this catalog. In Fig. \ref{fig:ssfrsm_cluster}, we again compare the SSFR--\sm~ relation against field galaxies but for 137 PFOF clusters selected with richness $> 25$, roughly corresponding to $M_{halo} > 10^{14}$ \Msolar. We found that the SSFR for star-forming galaxies with \sm~$ > 10^{9}$ \Msolar~ in the clusters, in contrast to group galaxies, is lower than field star-forming galaxies by $17\%$ with 4$\sigma$ confidence. Furthermore, the environment quenching efficiency, is in general higher in the cluster case than in the group, as revealed in Fig. \ref{fig:fq_cluster} (see Fig. \ref{fig:fq} for the group results). 

The higher quiescent fraction seen in the clusters compared to the field may be a result of the global reduction of the SSFR in the cluster galaxies, and/or proportionally more galaxies are being quenched in the groups. To examine the relative importance between the two effects, we also compute the fraction of galaxies below the SSFR threshold value of $10^{-10}\,$yr$^{-1}$ in the case where the SSFR of field galaxies are lower by the amount of SSFR difference seen between the field and clusters. The results are shown in green colors in Fig. \ref{fig:fq_cluster}. It reveals that the moderate reduction of the SSFR of the star-forming sequence alone can not fully account for the excess of quiescent galaxies in the clusters. This implies that the influence of cluster environment may consists of both slow and fast quenching mechanisms: the former is responsible for the gradual suppression of the SFR, while the latter must happen in a timescale short enough to produce a notable difference in the quiescent fraction between the field and clusters. We will discuss this issue more in \S \ref{sec:discussion}.

Now that we have shown that both the mass quenching efficiency $\epsilon^{mass}$ (or expressed as $f_{q}^{field}$) and the environment quenching efficiency $\epsilon^{envi}$ increase with the stellar mass, we proceed to compare the relative roles of these two effects as a function of stellar mass. The values of $\epsilon^{mass}$ and $\epsilon^{envi}$ are given in Table \ref{tab:qe}. In the bottom panel of Fig. \ref{fig:fq} and Fig. \ref{fig:fq_cluster}, we overplot $\epsilon^{mass}$ (same as the quantities $f_{q}^{field}$ shown on the top panel). In the lower redshift bin ($0.2 < z < 0.5$), it is clear that the quenching process is dominated by the  mass quenching for more massive galaxies whereas the environment is a secondary effect. However, the environment quenching exceeds the mass quenching for galaxies with stellar mass lower than a  $1-2\times10^{10}$\Msolar. This transition mass increases slightly from the group to the cluster environments. In the higher redshift bin ($0.5 < z < 0.8$), there is also evidence that the mass quenching plays a more important role for massive galaxies, but we can not pinpoint the transition mass because the sample becomes incomplete below $3\times10^{10}$\Msolar. 

\begin{deluxetable*}{lcccc}
\tabletypesize{\scriptsize}
\tablewidth{0pt}
\tablecaption{The mean SSFR and quiescent fraction in different environments.\label{tab:ssfrfq}}
\tablehead{
    \colhead{Subsample} &
    \colhead{Stellar mass} &
    \colhead{SSFR $^{ALL}$ (yr$^{-1}$)} &
    \colhead{SSFR $^{SF}$ (yr$^{-1}$)} &
    \colhead{$f_{q}$}
}

\startdata
Field   ($0.2 < z < 0.5$)           & 9.24 & $-9.23 \pm 0.00$ & $-9.17 \pm 0.00$ & $0.05 \pm 0.00$ \\
            & 9.55 & $-9.35 \pm 0.01$ & $-9.23 \pm 0.00$ & $0.10 \pm 0.00$ \\
            & 9.84 & $-9.51 \pm 0.01$ & $-9.31 \pm 0.01$ & $0.18 \pm 0.01$ \\
            & 10.14 & $-9.73 \pm 0.01$ & $-9.43 \pm 0.01$ & $0.31 \pm 0.01$ \\
            & 10.44 & $-10.00 \pm 0.01$ & $-9.57 \pm 0.01$ & $0.53 \pm 0.02$ \\
            & 10.74 & $-10.20 \pm 0.01$ & $-9.71 \pm 0.01$ & $0.70 \pm 0.02$ \\
            & 11.04 & $-10.30 \pm 0.01$ & $-9.71 \pm 0.03$ & $0.84 \pm 0.04$ \\
            & 11.33 & $-10.23 \pm 0.02$ & $-9.74 \pm 0.03$ & $0.70 \pm 0.05$ \\
Group   ($0.2 < z < 0.5$)           & 9.25 & $-9.24 \pm 0.01$ & $-9.15 \pm 0.01$ & $0.08 \pm 0.01$ \\
            & 9.55 & $-9.51 \pm 0.02$ & $-9.29 \pm 0.01$ & $0.21 \pm 0.02$ \\
            & 9.85 & $-9.74 \pm 0.02$ & $-9.34 \pm 0.02$ & $0.36 \pm 0.03$ \\
            & 10.15 & $-10.01 \pm 0.02$ & $-9.50 \pm 0.02$ & $0.52 \pm 0.03$ \\
            & 10.45 & $-10.15 \pm 0.02$ & $-9.58 \pm 0.02$ & $0.70 \pm 0.04$ \\
            & 10.73 & $-10.31 \pm 0.01$ & $-9.73 \pm 0.02$ & $0.83 \pm 0.05$ \\
            & 11.05 & $-10.37 \pm 0.02$ & $-9.84 \pm 0.02$ & $0.87 \pm 0.07$ \\
            & 11.34 & $-10.40 \pm 0.03$ & $-9.90 \pm 0.02$ & $0.93 \pm 0.12$ \\
Cluster ($0.2 < z < 0.5$)           & 9.25 & $-9.38 \pm 0.03$ & $-9.22 \pm 0.03$ & $0.14 \pm 0.03$ \\
            & 9.55 & $-9.62 \pm 0.04$ & $-9.31 \pm 0.03$ & $0.26 \pm 0.04$ \\
            & 9.84 & $-9.90 \pm 0.04$ & $-9.38 \pm 0.03$ & $0.50 \pm 0.06$ \\
            & 10.14 & $-10.12 \pm 0.03$ & $-9.47 \pm 0.03$ & $0.67 \pm 0.06$ \\
            & 10.45 & $-10.24 \pm 0.03$ & $-9.64 \pm 0.03$ & $0.77 \pm 0.08$ \\
            & 10.73 & $-10.29 \pm 0.03$ & $-9.74 \pm 0.04$ & $0.81 \pm 0.09$ \\
            & 11.05 & $-10.36 \pm 0.03$ & $-9.70 \pm 0.12$ & $0.90 \pm 0.13$ \\
            & 11.33 & $-10.42 \pm 0.04$ & $-9.95 \pm 0.05$ & $0.92 \pm 0.20$ \\
Field   ($0.5 < z < 0.8$)           & 9.25 & $-8.99 \pm 0.00$ & $-8.96 \pm 0.00$ & $0.02 \pm 0.00$ \\
            & 9.55 & $-9.13 \pm 0.00$ & $-9.08 \pm 0.00$ & $0.03 \pm 0.00$ \\
            & 9.84 & $-9.26 \pm 0.00$ & $-9.19 \pm 0.00$ & $0.05 \pm 0.00$ \\
            & 10.14 & $-9.47 \pm 0.01$ & $-9.32 \pm 0.00$ & $0.13 \pm 0.00$ \\
            & 10.44 & $-9.80 \pm 0.01$ & $-9.45 \pm 0.00$ & $0.33 \pm 0.01$ \\
            & 10.74 & $-10.14 \pm 0.01$ & $-9.58 \pm 0.01$ & $0.56 \pm 0.01$ \\
            & 11.04 & $-10.39 \pm 0.01$ & $-9.67 \pm 0.01$ & $0.77 \pm 0.02$ \\
            & 11.34 & $-10.58 \pm 0.01$ & $-9.72 \pm 0.02$ & $0.89 \pm 0.03$ \\
Group   ($0.5 < z < 0.8$)           & 9.25 & $-9.02 \pm 0.02$ & $-8.96 \pm 0.01$ & $0.04 \pm 0.01$ \\
            & 9.56 & $-9.15 \pm 0.02$ & $-9.09 \pm 0.01$ & $0.04 \pm 0.01$ \\
            & 9.86 & $-9.33 \pm 0.02$ & $-9.22 \pm 0.01$ & $0.08 \pm 0.01$ \\
            & 10.14 & $-9.60 \pm 0.02$ & $-9.39 \pm 0.01$ & $0.19 \pm 0.02$ \\
            & 10.45 & $-9.91 \pm 0.03$ & $-9.51 \pm 0.02$ & $0.40 \pm 0.03$ \\
            & 10.74 & $-10.18 \pm 0.02$ & $-9.61 \pm 0.02$ & $0.60 \pm 0.04$ \\
            & 11.06 & $-10.44 \pm 0.02$ & $-9.70 \pm 0.02$ & $0.83 \pm 0.06$ \\
            & 11.34 & $-10.57 \pm 0.02$ & $-9.74 \pm 0.04$ & $0.93 \pm 0.08$ \\
Cluster ($0.5 < z < 0.8$)           & 9.24 & $-9.02 \pm 0.04$ & $-8.95 \pm 0.03$ & $0.05 \pm 0.02$ \\
            & 9.55 & $-9.29 \pm 0.05$ & $-9.15 \pm 0.03$ & $0.10 \pm 0.03$ \\
            & 9.86 & $-9.42 \pm 0.05$ & $-9.26 \pm 0.03$ & $0.13 \pm 0.04$ \\
            & 10.15 & $-9.70 \pm 0.05$ & $-9.44 \pm 0.03$ & $0.26 \pm 0.05$ \\
            & 10.45 & $-10.05 \pm 0.06$ & $-9.53 \pm 0.03$ & $0.49 \pm 0.08$ \\
            & 10.75 & $-10.29 \pm 0.05$ & $-9.70 \pm 0.03$ & $0.64 \pm 0.10$ \\
            & 11.06 & $-10.51 \pm 0.04$ & $-9.73 \pm 0.05$ & $0.89 \pm 0.12$ \\
            & 11.34 & $-10.51 \pm 0.05$ & $-9.78 \pm 0.06$ & $0.87 \pm 0.15$
\enddata

\end{deluxetable*}

\begin{deluxetable}{lcccc}
\tabletypesize{\scriptsize}
\tablewidth{0pt}
\tablecaption{The mass quenching efficiency ($\epsilon^{mass}$) and environment quenching efficieny ($\epsilon^{envi}$) in groups and clusters. \label{tab:qe}}
\tablehead{
    \colhead{Redshift} &
    \colhead{Stellar mass} &
    \colhead{$\epsilon^{mass}$} &
    \colhead{$\epsilon^{envi}$ (group)} &
    \colhead{$\epsilon^{envi}$ (cluster)}
}

\startdata
$0.2 < z < 0.5$     & 9.25 & $0.05 \pm 0.00$ & $0.03 \pm 0.01$ & $0.10 \pm 0.03$ \\
            & 9.55 & $0.10 \pm 0.00$ & $0.12 \pm 0.02$ & $0.18 \pm 0.04$ \\
            & 9.85 & $0.18 \pm 0.01$ & $0.22 \pm 0.03$ & $0.39 \pm 0.07$ \\
            & 10.15 & $0.31 \pm 0.01$ & $0.30 \pm 0.05$ & $0.52 \pm 0.10$ \\
            & 10.45 & $0.53 \pm 0.02$ & $0.36 \pm 0.09$ & $0.52 \pm 0.16$ \\
            & 10.75 & $0.70 \pm 0.02$ & $0.44 \pm 0.18$ & $0.35 \pm 0.30$ \\
            & 11.05 & $0.84 \pm 0.04$ & $0.18 \pm 0.52$ & $0.36 \pm 0.87$ \\
            & 11.35 & $0.70 \pm 0.05$ & $0.77 \pm 0.43$ & $0.73 \pm 0.70$ \\
$0.5 < z < 0.8$     & 9.25 & $0.02 \pm 0.00$ & $0.01 \pm 0.01$ & $0.03 \pm 0.02$ \\
            & 9.55 & $0.03 \pm 0.00$ & $0.01 \pm 0.01$ & $0.07 \pm 0.03$ \\
            & 9.85 & $0.05 \pm 0.00$ & $0.03 \pm 0.01$ & $0.09 \pm 0.04$ \\
            & 10.15 & $0.13 \pm 0.00$ & $0.08 \pm 0.02$ & $0.15 \pm 0.06$ \\
            & 10.45 & $0.33 \pm 0.01$ & $0.10 \pm 0.05$ & $0.24 \pm 0.12$ \\
            & 10.75 & $0.56 \pm 0.01$ & $0.07 \pm 0.10$ & $0.18 \pm 0.23$ \\
            & 11.05 & $0.77 \pm 0.02$ & $0.25 \pm 0.26$ & $0.51 \pm 0.56$ \\
            & 11.35 & $0.89 \pm 0.03$ & $0.34 \pm 0.83$ & $-0.26 \pm -1.49$
\enddata

\end{deluxetable}
\section{DISCUSSION}\label{sec:discussion}

Our results suggest that there are three effects contributing to the
observed SFR -- density relation: firstly, the quiescent fraction
increases with the stellar mass; secondly, galaxies are progressively
more massive in the group environment; and last, at a given stellar mass, the quiescent fraction is higher in the group environment than in the field. Furthermore, the effect of being in the group environment is primarily to increase the fraction of the quiescent population, rather than reducing the SFR of the entire population globally as the star formation activity of the star-forming galaxies in groups is not distinctly different from those in the field (Fig. \ref{fig:ssfrsm}). Our main finding of this work is in good agreement with many previous studies \citep{bal04, vul10,li11,wij12,koy13}. However, our sample size is by far the largest among similar studies at the intermediate redshifts, and our sample includes both star-forming and passive populations that allow us to probe not only the properties of main-sequence galaxies but also the relative fraction between the star-forming and quiescent populations.

It has been long suggested that the bimodality of the galaxy colors \citep{bla03} implies a rapid star formation quenching process in moving galaxies from the blue cloud to the red sequence. The result that the SFR of star-forming sequence in the group environment is not globally reduced compared to their counterparts in the field suggests that the quenching mechanism operating in the groups/clusters is also a process during which the star formation is truncated in a very short period of time, leaving the SFR--\sm~ relation unaffected while increasing the quiescent population. In the absence of strong supernova feedback, this would favor the galaxy merger and ram-pressure stripping (of cold disk gas) scenarios which act on a short timescale $< 1$~Gyr \citep{gun72,lot10,jia12}, over other mechanisms often invoked to explain the origin of the observed morphology--density or color--density relations found in clusters, such as strangulation, galaxy harassment, etc. However, ram-pressure stripping is thought to be more effective for clusters with mass $>$ 10$^{14}$\Msolar~\citep{qui00,bek09}. It is unlikely that it dominates the trend seen in our group sample, and therefore this leaves mergers as the most plausible process transforming galaxies from the star-forming sequence to the quiescent populations. Interestingly, earlier works based on the galaxy luminosity functions of blue and red galaxies in groups also infer that the changes in the galaxy properties in galaxy groups are consistent with the effects due to galaxy interactions \citep{rob06,rob10}. Recent environment studies of galaxy mergers found that while most of the mergers occur in the intermediate environment which dominates the density distribution, the chance of galaxy mergers is actually higher in denser environments \citep{bal04,lin10,kam13}, peaked at the group environment \citep{jia12}. This provides another piece of evidence supporting the importance of mergers in high-density environments. 

In the absence of strong feedback, all the above arguments lead to the conclusion that galaxy mergers are a favored process of environment quenching acting on group galaxies. Nevertheless, more recent theories of galaxy formation invoke very strong feedback to explain the low efficiency of star formation \citep{opp08,whi91,bal04,bow06,opp08,bow12}. In these models, galaxies
receive gas flows into galaxies from the surrounding sheets and filaments at a high rate, but most of the incoming material is expelled by the galactic wind rather than formed into stars. When the galaxy becomes a satellite, this inflow is cut-off but the strong outflow continues to remove material from the galaxy. Because of the resulting imbalance between inflow and outflow, strangulation causes a rapid decline of the overall star formation rate in these models \citep{fon08}. In these theories, strangulation also provides a viable explanation for the rapid 
transformation that is seen in the data \citep{mcg09,wet13}.

When we compare the SFR and \sm~ distribution between field and cluster galaxies, we find that not only the quiescent fraction is significantly higher in the cluster environment, a moderate difference (17\%) in the star formation activities between cluster and field SF galaxies for a fixed stellar mass is also detected. The suppression of the SFR of star-forming galaxies in clusters is possibly due to a longer-time scale quenching process (eg., strangulation and galaxy harassment). However, we can not rule out the possibility that cluster members have been formed earlier and evolved over a longer period of time as opposed to field galaxies, and as a consequence have older stellar ages. Nevertheless, as shown in Fig. \ref{fig:fq_cluster}, the amount of depleted SSFR in clusters can not account for the excess of the quiescent fraction relative to field, implying that a rapid quenching process is relatively more efficient than the mechanism that causes the suppression in the SSFR of star-forming cluster galaxies.

We can also compare our results with previous studies of cluster samples in a similar redshift range. \citet{muz12} has studied the SSFR--\sm~ relation and the star-forming fraction for a sample of 9 clusters at $z \sim 1$ taken from the Gemini Cluster Astrophysics Spectroscopic Survey (GLASS). They found that at a fixed stellar mass, the fraction of star-forming galaxies strongly correlates with the environment, but the SSFR--\sm~relation is nearly identical between the field and clusters. A recent work by \citet{koy13} investigated the star formation properties of $H_{\alpha}-$selected galaxies for 3 clusters at $z \sim$ 0.4, 0.8, and 2.2 respectively and reached a similar conclusion that there is no notable difference in the SSFR--\sm~ relation between different environments. In contrast, \citet{vul10} detected a lower SSFR for star-forming members in clusters than their field counterparts by a factor of 1.5 based on 16 clusters at $0.4 < z < 0.8$ drawn from the Distant Cluster Survey (EDisCS). Most recently, \citet{zei13} also found that the SSFR of star-forming galaxies for a sample of 18 clusters at $1 < z < 1.5$ drawn from the IRAC Shallow Cluster Survey is systematically lower than the field galaxies at a fixed stellar mass. The discrepancy between these results is possibly due to the small size of the cluster samples utilized in their analysis. As revealed from our studies, the difference in the SSFR at a fixed stellar mass between the
field and cluster galaxies is only moderate (17\%), and therefore it is likely that the degree of SSFR suppression is under- or over-estimated in previous studies due to the small number statistics.

Next we turn into the discussion of other implications of the
similarity in the SFR--\sm~ relation between different
environments. The star formation activity is believed to be regulated
by the fueling of fresh gas and the feedback processes. Several galaxy
formation models have shown that the star formation rates are strongly
related to the gas accretion rates, and the tightness of the
SFR--\sm~relation is a natural outcome of the relatively smooth and
steady gas accretion of haloes \citep{bow06,dav08,dut10}. In these
models, the characteristics of the SFR--\sm~ relation is insensitive
to the details in the feedback models, but governed by the rate of gas
accretion, and hence by the rate of halo growth. If it is indeed true
that there is a strong relationship between the SFR and the cold-gas
accretion rate, one might naively expect that the SFR can differ in
different environments as the gas accretion properties are dependent
on the location of the cosmic web and on the halo mass. This leads to
a puzzle as to why,  in contrast to the simple expectation,
the observed SFR is at a similar level for
star-forming group galaxies as it is for field counterparts selected
at the same stellar mass.
A possible explanation is that the star forming galaxies have only recently entered the group environment and are still embedded in distinct filaments or sub-haloes that maintain the supply of gas onto the central galaxy. However, in order to make this viable the delay during which the gas supply is maintained must be long compared to the orbital time and must scale with redshift \citep{mcg09,wet13,mok13}.

One of the interesting results from this work is the suggestion that the environment quenching efficiency is stronger for more massive galaxies. This is in contradiction to a naive expectation that smaller galaxies are more vulnerable to effects that remove gas supplies, such as tidal disruption or ram pressure stripping, and therefore have a stronger SFR suppression or quenching effect. Such a trend might result if more massive objects tend to be accreted on more radial objects, or experience more dramatic angular momentum loss
due to dynamical friction, so that the haloes of these objects experience greater ram pressure. Alternatively, it may simply arise from the biases in halo formation histories, with more massive galaxies being more likely to have been ``pre-processed'' in smaller groups prior to their accretion. On the other hand, there is also evidence showing that galaxy merger rate increases with galaxy luminosity (and hence stellar mass) \citep{pat08}, which might be able to explain the increased quenching efficiency with increasing stellar mass in the groups. Nevertheless, it is worth noting that the mass limit of our sample is still constrained by the current MD depth. To push the analysis down to smaller mass range for more robust conclusions have to await till the completion of the full PS1/MD surveys. 

Interestingly the mass-dependent environment quenching is also apparent in the strangulation model introduced by \citep{fon08}. By comparing the passive fractions in different bins of halo mass, we computed the environment quenching efficiency in the model in the same way we did for the observed data (i.e., using Eq. \ref{eq:epsilon}). The results are shown as a function of stellar mass in Fig. \ref{fig:fq} and \ref{fig:fq_cluster}. It is evident that $\epsilon^{envi}$ increases with stellar mass in this model, similarly to the observed trend. We emphasize that this stellar mass dependence of the environment quenching efficiency is not built in by hand but is rather a consequence of the Font et al. model. Therefore this provides a further support of the strangulation process adopted in \citet{fon08} being a plausible mechanism that is responsible for the environment quenching in partcular in the clusters.

Whether or not deeper data substantiate the suggestion that the impact
of environment is greater for more massive galaxies, the relative
importance of the environment with respective to mass quenching declines with the stellar mass as
revealed in Fig. \ref{fig:fq}, and Fig. \ref{fig:fq_cluster}. The
transition mass in which the quenching in central galaxies dominates
is at roughly $1-2\times10^{10}$\Msolar, in good agreement with previous
work \citep{kau03,pen10}. Our results that the environment quenching
is more important for small galaxies may shed light on the cosmic
evolution of the galaxy number densities. For example, previous works
have found that the number density of red galaxies has increased by at
least a factor of two since $z\sim1$ while the number density of blue
galaxies has remained roughly constant \citep{bel04,fab07}. Using a
stellar mass-selected sample drawn from the PRIMUS survey
\citep{coi11}, \citet{mou13} further found that the increase in the
number density of red galaxies is more prominent for low mass
galaxies. Qualitatively at least, this is consistent with the growth
of large-scale structure. As the large-scale structure builds up, the
number of galaxies belonging to the group and/or cluster environments
increases with time. As a result, the number of low-mass galaxies
being quenched through the environment effect also increases towards
lower redshift, leading to the buildup of the red sequence over cosmic
time.  The challenge of course, is to reproduce the quantitative details of this behaviour in cosmological models.

\section{CONCLUSION}
We have carried out an analysis of the SFR -- \sm~ relation between field and group galaxies to study the environmental effect on shaping galaxy properties out to $z \sim 0.8$ using data from the early PS1 Medium Deep survey. Galaxy groups are identified using the group-finder PFOF \citep{jia13} that is optimized for the PS1 photometric redshift sample. Galaxies are divided into the `star-forming' (SF) and `quiescent' (non-SF) populations by their specific star formation rate with a threshold of $10^{-10}$ per year. We study the SFR -- \sm~ relation with and without the inclusion of quiescent populations, and the quiescent fraction as a function of stellar mass in different environments. And finally we investigate the stellar mass dependence of the environment quenching efficiency.

 Our main conclusions are as follows:

1. Group galaxies are found to be systematically more massive than galaxies in the field for both star-forming and quiescent populations, supporting the hierarchical scenario that galaxies are formed earlier in denser environments. In this picture, the existence of massive star-forming group galaxies in the higher-redshift bin ($0.5 < z < 0.8$) is a consequence of an earlier and faster evolution in group environment.

2. The normalization and slope of the SSFR -- \sm~ relation for star-forming galaxies are comparable between the field and group environments. On the other hand, the normalization is different when the quiescent galaxies are included in the analysis, being reduced for group galaxies. 

3. Over the redshift range $0.2 < z <0.8$, we find that the quiescent
   fraction is a strong function of the stellar mass, and is greater
   in the groups than in the field. The SFR -- density relation is the
   combination of these two effects plus the result from point 1 above. The excess of the quiescent fraction in the groups is responsible for the difference of the SFR -- \sm~ relation (all galaxies included) seen between field and groups. 

4. The lack of the SFR suppression for star-forming galaxies and the
   higher quiescent fraction in groups suggest a fast quenching
   mechanism acting in the group environment. Galaxies, if being
   quenched, must have been moved from the star-forming sequence to
   the quiescent population in a relatively short time-scale to
   preserve the SFR -- \sm~ relation of the main-sequence. This favors
   galaxy mergers as a primary process that quenches the
   star-formation activities in galaxy groups, as other mechanisms are
   either inefficient at group scales (e.g., ram-pressure stripping) or
   operate over a longer time-scale (e.g., strangulation, and galaxy harassment, etc.) in the absence of strong feedback. 

5. In contrast to the group environment, clusters have a more
   prominent effect on reducing the SSFR of the star-forming
   sequence. For a given stellar mass, the SFR is moderately lower by
   17\% at the 4$\sigma$ confidence level in clusters than its field
   counterpart. Moreover, the quiescent fraction for clusters is found to be
   greater than that in both the group and field environments, leading to a
   greater environment quenching efficiency. The amount of the reduction in the SSFR of star-forming galaxies in clusters can not fully account for the difference in the quiescent fraction between clusters and the field, however. This implies that the quenching mechanisms acting on the clusters involve both fast and slow processes.

6. In both the field and group environments, the quiescent fraction is
   a steep function of the stellar mass, being higher in more massive
   systems. Our results also indicate that the environment quenching
   efficiency increases with stellar mass, albeit a large sample is
   needed to draw a firm conclusion. A similar trend is also visible
   in the \citet{fon08} model. At the cluster scale, this can be attributed to the fact that massive galaxies are subject to stronger dynamical friction and shrink into the central part of clusters in a shorter timescale where the effect of tidal disruption and the ram-pressure is strongest. In groups, it is likely due to the increased galaxy merger frequency with increasing stellar mass. 

7. The relative importance between mass quenching and environment quenching depends on the stellar mass of galaxies. In the lower redshift bin ($0.2 < z <0.5$), the mass quenching plays a dominant role in producing quiescent galaxies for more massive galaxies, while less massive galaxies are quenched mostly through the environmental effect. This transition mass is around $1-2\times10^{10}$\Msolar~ in group and cluster environments.

\acknowledgments

We thank the anonymous referee for his/her constructive suggestions that further improve this paper. We also thank M. Kriek for releasing her SED fitting code `FAST', and O. Ilbert and the COSMOS team for the usage of the stellar mass measurement from the unpublic V1.7 photometric redshift catalog.  L. Lin thanks F. Van de Voort, J. Heasley, and Y.-T. Lin for very helpful discussions, and N. Mostek for advising how their methods of estimating star formation rate can be used. We also thank 
PS1 staff who have provided help with the PS1 data observation and reduction. The work is supported by the National Science Council of Taiwan
under the grant NSC99-2112-M-001-003-MY3 and NSC 101-2112-M-001-011-MY2. H.-Y. Jian acknowledges the support of NSC101-2811-M-002-075. W.-P. Chen acknowledges the support of NSC 102-2119-M-008-001. P. Norberg acknowledges the support of the Royal Society through the award of a University Research Fellowship and the
European Research Council, through receipt of a Starting Grant (DEGAS-259586).
PAM acknowledges support from a European Research Council Starting Grant (DEGAS-259586). The Pan-STARRS1 Surveys (PS1) have been made possible through contributions of the Institute for Astronomy, the University of Hawaii, the Pan-STARRS Project Office, the Max-Planck Society and its participating institutes, the Max Planck Institute for Astronomy, Heidelberg and the Max Planck Institute for Extraterrestrial Physics, Garching, The Johns Hopkins University, Durham University, the University of Edinburgh, Queen's University Belfast, the Harvard-Smithsonian Center for Astrophysics, the Las Cumbres Observatory Global Telescope Network Incorporated, the National Central University of Taiwan, the Space Telescope Science Institute, the National Aeronautics and Space Administration under Grant No. NNX08AR22G issued through the Planetary Science Division of the NASA Science Mission Directorate, the National Science Foundation under Grant No. AST-1238877, the University of Maryland, and Eotvos Lorand University (ELTE). We close with thanks to Hawaiian people for the use of their sacred mountain.

\appendix
\section{Field Contamination and Group Completeness Corrections} \label{app:corr}
In this section, we describe the method we use to compute the field
contamination rate $R_{c}$ and the group member recovery rate $R_{r}$
that are used to account for the contamination and incompleteness
effects introduced in the PFOF group identification. We first cross
match the MD07 photometric redshift catalog with the DEEP2 EGS
spectroscopic galaxy sample \citep{new13}. We then define a subset of
the sample which only contains galaxies with secure spectroscopic redshift measurements. For each galaxy in this subsample, we have two flags indicating whether they are group members based on 1) MD PFOF's identification used in this work and 2) the DEEP2 group identification by \citet{ger12}. $R_{r}$ is then calculated as the fraction of DEEP2 spectroscopically-identified group members that are also PFOF members, while $R_{c}$ is computed as the fraction of PFOF group members that are not associated with  spectroscopically-identified groups. 

\begin{figure}[h]

\includegraphics[angle=0,width=17cm]{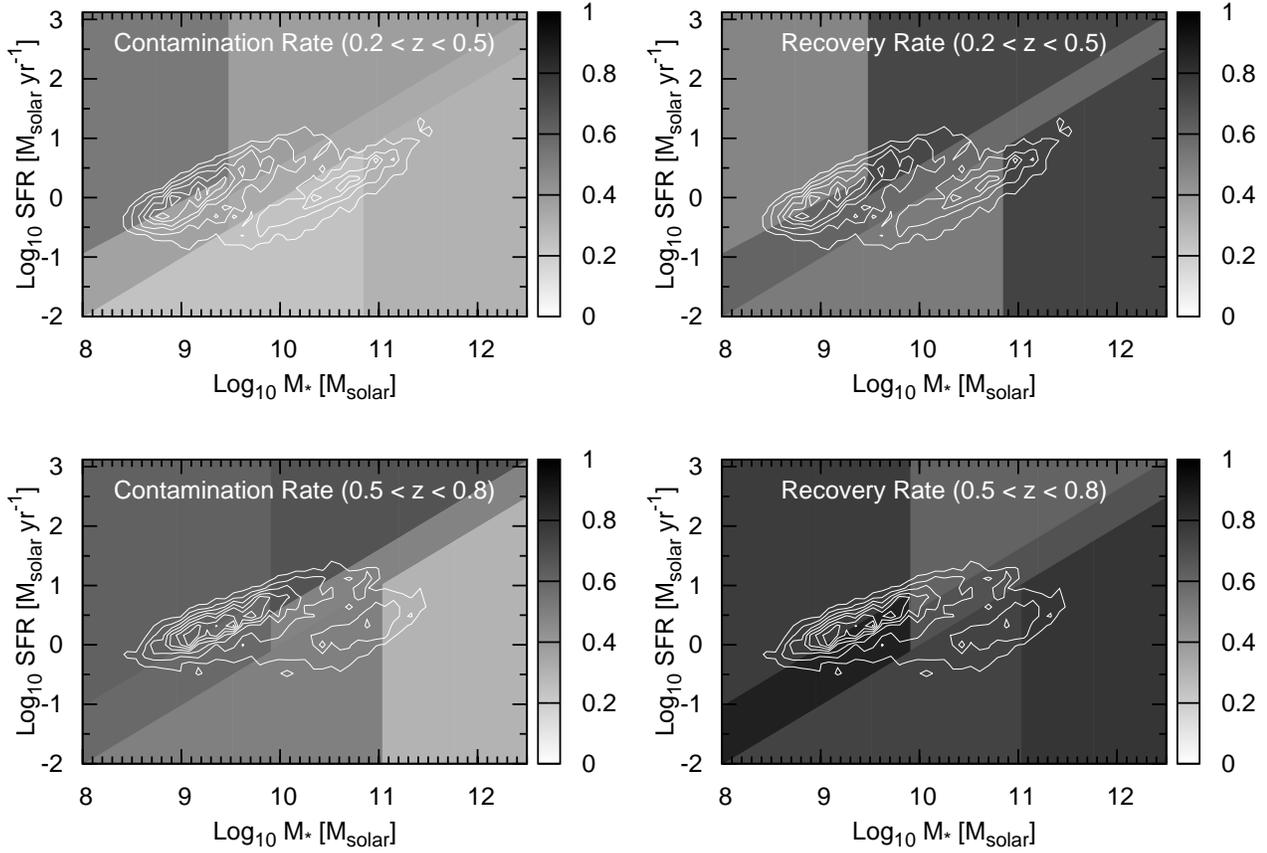}

\caption{The correction factors for PFOF-identified groups. The field contamination rate $R_{c}$ (left panels) and the group member recovery rate $R_{r}$ (right panels) are shown in the background. The white contours show the density distribution of PFOF group galaxies. 
\label{fig:corr}}
\end{figure}

Since we are interested in the properties of galaxies on the SFR and
\sm~ plane, ideally the two quantities $R_{c}$ and $R_{r}$ should be
computed on a fine grid of SFR and \sm. However, the correction scheme
we adopt relies on the spectroscopic sample whose sample size is not
large enough for this purpose. We instead divide the galaxies into six sub-regions on the SFR and \sm~ plane when computing $R_{c}$ and $R_{r}$. SF galaxies are grouped into four main regions: above and below the main sequence, for both large and small galaxies; while the quiescent galaxies are divided into large and small galaxies. For galaxies that are less massive than the mass limit of the spectroscopic redshift sample due to the difference in the depth and filter cut between the two surveys ($R < 24.1$ for the EGS vs $i_{\mathrm{P}1} < 24$ for the PS1 MD), we adopt the same values of $R_{c}$ and $R_{r}$ as those derived in the smallest mass bins. The resulting $R_{c}$ and $R_{r}$ are shown in Fig. \ref{fig:corr}. To test the robustness of our correction factors, we have also tested our results by varying the number of sub-regions and the choice of the dividing lines, and found that none of our conclusions presented in this work is significantly changed.

\end{document}